\documentclass[openany,USenglish]{article}

\usepackage[utf8]{inputenc}
\usepackage{microtype}

\usepackage[section]{algorithm}
\usepackage{algpseudocode}

\usepackage{color}

\usepackage{mathrsfs}

\usepackage{amsmath}
\usepackage{amssymb}
\usepackage{graphicx}
\usepackage{hyperref}

\newcommand{\vfabs}[1]{\left \lvert #1 \right\rvert }
\newcommand{\vfnorm}[1]{ \left \lVert #1 \right \rVert} 
\newcommand{\vfRe}[1]{\mathrm{R}\mathrm{e}\left \{#1\right\} }
\newcommand{\vfIm}[1]{\mathrm{I}\mathrm{m}\left \{#1\right\} }

\begin{document}



\author{Piero Triverio\footnote{Edward S. Rogers Sr. Department of Electrical \& Computer Engineering, University of Toronto, 10 King's College Rd.,
	M5S 3G4, Toronto, ON, Canada, e-mail: piero.triverio@utoronto.ca}}
\title{Vector Fitting}

\maketitle

\section*{Abstract}
	We introduce the Vector Fitting algorithm for the creation of reduced-order models from the sampled response of a linear time-invariant system. This data-driven approach to reduction is particularly useful when the system under modeling is known only through experimental measurements. The theory behind Vector Fitting is presented for single- and multiple-input systems, together with numerical details, pseudocodes, and an open-source implementation. We discuss how the reduced model can be made stable and converted to a variety of forms for use in virtually any modeling context. Finally, we survey recent extensions of the Vector Fitting algorithm geared towards time-domain, parametric and distributed systems modeling. This work is a draft of the book chapter P. Triverio, ``Vector Fitting'' that will be part of the ``Handbook on Model Order Reduction'' edited by P. Benner, S. Grivet-Talocia, A. Quarteroni, G. Rozza, W. H. A. Schilders, L. M. Silveira, to appear for De Gruyter.


\section{Introduction and motivation}

The Vector Fitting (VF) algorithm~\cite{gustavsen1999rational,grivet2015passive} is one of the most successful techniques for creating reduced-order models for linear systems starting from \emph{samples} of their response. Samples may originate from an experimental measurement or from a prior numerical simulation.  This need arises in many practical scenarios, and we cite two examples. 

A biomedical engineer may need a linear model describing blood flow in a portion of the human cardiovascular system, and have simultaneous in-vivo measurements of pressure and flow rate at the inlets and outlets of the region of interest. With a \emph{data-driven} algorithm for model order reduction, such as VF, the reduced model can be created directly from experimental observations. 

As a second example, we consider an electronic engineer that needs a model for a Radio-Frequency (RF) amplifier or an antenna, to be used for design purposes. If the device is provided by a third party, a measurement may be the only way to characterize the system. High-frequency measurements are typically performed in the frequency domain, and return the impedance or admittance seen between the ports of the device, measured at various frequencies $\omega_k$. From these samples, VF can create a reduced model which can be represented as a set of differential equations or as an equivalent circuit for use in subsequent simulation, including those performed in the time domain.

The main advantage of a data-driven approach to reduced-order modeling is that only samples of the system response are required. This feature makes data-driven reduction a natural choice when experimental measurements are readily available. Furthermore, data-driven reduction can also be applied when samples originate from a numerical simulation based on first-principles equations, such as Maxwell's equations for electromagnetic phenomena. Although, in this second scenario, one could technically use \emph{equation-driven} methods, the available simulator may not allow the user to export the discretized first-principle equations for reduction. This is the case for most commercial simulators used by industry. The main disadvantage of data-driven reduction is that it offers less physical insight into the system under modeling, since it leads to a ``black-box'' reduced model. By starting from a first-principle model, equation-driven methods are typically better in this regard, since they can provide to the user  more information about which features of the original model were retained, and which features were discarded.

\section{The  Sanathanan-Koerner algorithm}
\label{sec:vf_sk}

\subsection{Problem statement}

We assume that the system under modeling is linear and time-invariant, with input $u(t) \in \mathbb{R}^{\bar{m}}$ and output $y(t) \in \mathbb{R}^{\bar{q}}$. Because of linearity and time-invariance, the output can be written as a convolution
\begin{equation}
	y(t) = \int_{-\infty}^{+\infty} h(t-\tau) u(\tau) d\tau
	\label{eq:vf_systd}
\end{equation}
between input $u(t)$ and the impulse response $h(t) \in \mathbb{R} ^{\bar{q} \times \bar{m}}$ of the system, which is unknown. Applying the Laplace transform to both sides of~\eqref{eq:vf_systd}, we get
\begin{equation}
	Y(s) = H(s) U(s)\,,
	\label{eq:vf_sysfd}
\end{equation}
where $s = \sigma + \jmath \omega$ is complex frequency. In~\eqref{eq:vf_sysfd}, $U(s) \in \mathbb{C}^{\bar{m}}$ and $Y(s) \in \mathbb{C}^{\bar{q}}$ are the Laplace transforms of $u(t)$ and $y(t)$, respectively, and $H(s) \in \mathbb{C}^{\bar{q} \times \bar{m}}$ is the transfer function of the system. The VF algorithm solves the following problem. Given $\bar{k}$ measurements of the transfer function
\begin{equation}
	H_k = H(\jmath \omega_k) \qquad k = 1,\dots, \bar{k}\,,
	\label{eq:vf_Hk}
\end{equation}
determine a rational function $\widetilde{H}(s)$ that approximates the given measurements
\begin{equation}
	\widetilde{H}(\jmath \omega_k) \simeq H_k \qquad \forall k = 1,\dots, \bar{k}\,.
	\label{eq:vf_fit}
\end{equation}
In VF, $\widetilde{H}(s)$ is chosen to be a rational function. Rational functions are universal approximators, and can therefore approximate a wide range of functions with arbitrary accuracy. Moreover, since the transfer function of lumped systems is rational by construction, this is a natural choice to model dynamical systems. Finally, rational functions can be represented as a state-space system, a poles-residue form, a set of differential equations, an equivalent electric circuit and many other forms. This flexibility facilitates the integration of the reduced model into existing software for computational mathematics and system simulation.

\subsection{The Levy and Sanathanan-Koerner algorithms}

The first attempts to solve~\eqref{eq:vf_fit} numerically date back at least to the 1950s, with the works of Levy, Sanathanan and Koerner among others. We briefly summarize their work since the VF algorithm can be better understood from that perspective. For simplicity, we initially consider the case of a system with a single input and a single output ($\bar{m}=\bar{q}=1$). The general case will be discussed in Sec.~\ref{sec:vf_mimo}. 

In order to solve the approximation problem~\eqref{eq:vf_fit}, we must first choose a suitable parametric form for $\widetilde{H}(s)$, which is the model that we want to estimate from the given samples. The most natural choice is to let $\widetilde{H}(s)$ be the ratio of two polynomials
\begin{equation}
	\widetilde{H}(s) =
	\frac{n(s)}{d(s)} = 
	 \frac{
		\sum_{n=0}^{\bar{n}} a_n s^n
	}{
		\sum_{n=0}^{\bar{n}} b_n s^n
	}\,,
	\label{eq:vf_Hsk}
\end{equation}
where $a_n, b_n \in \mathbb{R}$ are unknowns, and $\bar{n}$ is the order of the desired model. Since one coefficient can be normalized, we let $b_{\bar{n}} = 1$. In~\eqref{eq:vf_Hsk}, we chose the same degree $\bar{n}$ for numerator and denominator. This choice is appropriate for transfer functions that are known to be bounded when $s \to \infty$. This is the case of the scattering coefficients used to model  electronic devices at high frequencies, as in the example in Sec.~\ref{sec:vf_ex3}. In other applications, the transfer function of the system under modeling may  grow polynomially as $s$ increases. This is the case, for example, of the impedance and admittance coefficients of passive electrical circuits, which can grow linearly with $s$. As an example, one can consider the impedance $Z(s) = s L$ of an inductor. In such cases, the degree of the numerator of~\eqref{eq:vf_Hsk} should be increased to $\bar{n}+1$. This change leads to minor modifications to the algorithms  presented in this chapter, which will not be discussed here, but can be found in~\cite{grivet2015passive}.

After choosing the form of model~\eqref{eq:vf_Hsk} , we have to determine its coefficients $a_n$ and $b_n$ in order to satisfy~\eqref{eq:vf_fit}, minimizing a suitable norm between samples $H_k$ and model response $\widetilde{H}(\jmath \omega_k)$. We choose the $l_2$ norm, and aim to minimize 
\begin{equation}
	e^2 = \frac{1}{\bar{k}} \sum_{k=1}^{\bar{k}} \left|	
		H_k - \widetilde{H}(\jmath \omega_k)
	\right|^2\,.
	\label{eq:vf_e}
\end{equation}
Minimizing~\eqref{eq:vf_e} is a nonlinear least squares problem, due to the unknowns $b_n$ in the denominator. Although nonlinear optimization algorithms can be directly applied to~\eqref{eq:vf_e}, experience shows that they can be quite time consuming and prone to local minima. A different approach is preferred, where~\eqref{eq:vf_e} is linearized into a \emph{linear} least squares problem, which can be solved efficiently and robustly with the QR decomposition~\cite{Gol96}.

We first rewrite~\eqref{eq:vf_e} as
\begin{equation}
	e^2 = \frac{1}{\bar{k}}  \sum_{k=1}^{\bar{k}} \left|	
	 \frac{ H_k \sum_{n=0}^{\bar{n}} b_n (\jmath \omega_k)^n -  \sum_{n=0}^{\bar{n}} a_n (\jmath \omega_k)^n}{\sum_{n=0}^{\bar{n}} b_n (\jmath \omega_k)^n}
	\right|^2\,.
	\label{eq:vf_e2}
\end{equation}
Levy proposed to linearize~\eqref{eq:vf_e2} by simply neglecting the denominator, and minimize~\cite{levy1959complex}
\begin{equation}
	\left(e_{L}\right)^2 = \frac{1}{\bar{k}}  \sum_{k=1}^{\bar{k}} \left|	
	H_k \sum_{n=0}^{\bar{n}} b_n (\jmath \omega_k)^n -  \sum_{n=0}^{\bar{n}} a_n (\jmath \omega_k)^n
	\right|^2\,,
	\label{eq:vf_eL}
\end{equation}
which ultimately boils down to solving a system of linear equations in least squares sense. Unfortunately, this simple trick typically fails to provide an accurate solution of~\eqref{eq:vf_fit}. Indeed, error functionals~\eqref{eq:vf_e2} and~\eqref{eq:vf_eL} are equivalent only when $\sum_{n=0}^{\bar{n}} b_n (\jmath \omega)^n$ is approximately constant, which is rarely the case. Furthermore, the monomial terms $(\jmath \omega)^n$ in~\eqref{eq:vf_eL} will result in Vandermonde matrices in the least squares problem to be solved, which are ill-conditioned~\cite{Gol96}.

To overcome this issue, Sanathanan and Koerner proposed an iterative process to improve the quality of the solution~\cite{sanathanan1963transfer}. In the first iteration ($i=1$), the Levy functional~\eqref{eq:vf_eL} is minimized, providing a first estimate of the model coefficients that we denote as $a_n^{(1)}$ and $b_n^{(1)}$. In successive iterations ($i \ge 2$), the following linearization of~\eqref{eq:vf_e2} is minimized
\begin{equation}
	\left( e_{SK}^{(i)} \right)^2 =
	 \frac{1}{\bar{k}}  \sum_{k=1}^{\bar{k}} \left|	
	\frac{ H_k \sum_{n=0}^{\bar{n}} b_n^{(i)} (\jmath \omega_k)^n -  \sum_{n=0}^{\bar{n}} a_n^{(i)} (\jmath \omega_k)^n}{\sum_{n=0}^{\bar{n}} b_n^{(i-1)} (\jmath \omega_k)^n}
	\right|^2\,,
	\label{eq:vf_eSK}
\end{equation}
leading to a new estimate of model coefficients $a_n^{(i)}$ and $b_n^{(i)}$. We can see that, in~\eqref{eq:vf_eSK}, the coefficients $b_n^{(i-1)}$ from the previous iteration are used to approximate the ``nonlinear'' term in~\eqref{eq:vf_e2}. Since unknowns $a_n^{(i)}$ and $b_n^{(i)}$ appear only in the numerator, the Sanathanan-Koerner method only requires the solution of linear least squares problems. If the iterative process converges,  $b_n^{(i-1)} \to b_n^{(i)}$, and~\eqref{eq:vf_eSK} becomes equivalent to~\eqref{eq:vf_e2}. We can see that the term $\sum_{n=0}^{\bar{n}} b_n^{(i-1)} (\jmath \omega_k)^n$ in the denominator of~\eqref{eq:vf_eSK} acts as a frequency-dependent weight of the least squares problem. This weight aims to progressively remove the bias introduced in the linearization of~\eqref{eq:vf_e}. For discrete-time systems, the counterpart of the Sanathanan-Koerner method was proposed by Steiglitz and McBride~\cite{steiglitz1965technique}.

\subsection{Numerical issues of the Sanathanan-Koerner method}
\label{sec:vf_nisk}

The work of Sanathanan and Koerner solves~\eqref{eq:vf_fit} accurately using only linear least squares problems. Unfortunately, this method can still suffer from severe numerical issues when applied to realistic problems, where the required model order $\bar{n}$ may be quite large and frequency $\omega$ may span several orders of magnitude. For example, VF is extensively used in integrated circuit design to model the interconnect network that distributes signals and power across the circuit. In this application, the frequency range of interest typically extends from a few MHz to tens of GHz, for about four decades of variation. The numerical issues associated with the Sanathanan-Koerner method arise from two factors:
\begin{enumerate}
	\item[a)] error~\eqref{eq:vf_eSK} contains high powers of frequency $(\omega_k)^n$, leading to very poor conditioning. Specifically, the matrix of the least-squares problem to be solve will contain Vandermonde blocks~\cite{Gol96}, which are known to be ill conditioned even for relatively modest values of $\bar{n}$;
	\item[b)] weighting term $\sum_{n=0}^{\bar{n}} b_n^{(i-1)} (\jmath \omega_k)^n $ in the denominator of~\eqref{eq:vf_eSK} typically exhibits large variations over $[\omega_1,\omega_{\bar{k}}]$, which further degrades the conditioning of the least squares problem.
\end{enumerate}

\section{The Vector Fitting algorithm}

The VF algorithm, conceived by Gustavsen and Semlyen~\cite{gustavsen1999rational}, addresses both problems with a simple yet brilliant solution. 

\subsection{A new basis function and implicit weighting}

In order to avoid the ill-conditioning arising from the $s^n$ terms in~\eqref{eq:vf_Hsk}, VF replaces those terms with partial fractions. The numerator and denominator of $\widetilde{H}(s)$ are written as
\begin{align}
	n^{(i)}  & =  c_0^{(i)} + \sum_{n=1}^{\bar{n}} \frac{c_n^{(i)}}{s-p_n^{(0)}}\,, 
		\label{eq:vf_n}\\
	d^{(i)}  & =  1 + \sum_{n=1}^{\bar{n}} \frac{d_n^{(i)}}{s-p_n^{(0)}}\,,
	\label{eq:vf_d}
\end{align}
where $p_n^{(0)} \in \mathbb{C}$ are a set of initial poles, whose choice will be discussed later on. We see that, without loss of generality, the constant term in~\eqref{eq:vf_d} has been normalized to one. In comparison to the monomial basis functions $s^n$ used by the Sanathanan-Koerner iteration, which vary wildly as $s$ increases, partial fractions $\frac{1}{s-p_n^{(0)}}$ have more contained variations over frequency if poles $p_n^{(0)}$ are chosen appropriately~\cite{hendrickx2006discussion}, as will be discussed in Sec.~\ref{sec:vf_siso}. This feature leads to better conditioning, especially if the poles $p_n^{(0)}$ are distinct and well separated.

The introduction of partial fractions is also crucial to address the second issue discussed in Sec.~\ref{sec:vf_nisk}, and perform an implicit weighting of~\eqref{eq:vf_eSK}. To understand how VF achieves this, we first give a different interpretation to linearized error~\eqref{eq:vf_eSK}. In terms of~\eqref{eq:vf_n} and~\eqref{eq:vf_d}, error~\eqref{eq:vf_eSK} can be expressed as
\begin{equation}
	\left( e_{SK}^{(i)} \right)^2 =
	\frac{1}{\bar{k}} \sum_{k=1}^{\bar{k}} \left|	
	H_k \frac{d^{(i)}(\jmath \omega_k)}{d^{(i-1)}(\jmath \omega_k)} - \frac{n^{(i)}(\jmath \omega_k)}{d^{(i-1)}(\jmath \omega_k)}
	\right|^2\,.
	\label{eq:vf_eSK2}
\end{equation}
We can see that this expression involves two new quantities
\begin{equation}
	w^{(i)}(s) 							 = \frac{d^{(i)}(s)}{d^{(i-1)}(s)}\,,
		\label{eq:vf_wbefore}
\end{equation}
and
\begin{equation}
	\widetilde{H}^{(i)}(s)	 = \frac{n^{(i)}(s)}{d^{(i-1)}(s)}\,.
			\label{eq:vf_Hbefore}
\end{equation}
Function $\widetilde{H}^{(i)}(s)$ can be interpreted as the model transfer function estimated at iteration $i$ by the minimization of~\eqref{eq:vf_eSK2}. Notably, this transfer function is made by the numerator $n^{(i)}(s)$ from the current iteration (to be found), and by the denominator $d^{(i-1)}(s)$ from the previous iteration (already known). This approximation arises from the linearization of the error function, since it indeed avoids the presence of unknowns in the denominator. Function  $w^{(i)}(s)$ can be interpreted as a frequency-dependent weight which multiplies the given samples $H_k$. This weighting function has two purposes:
\begin{itemize}
	\item providing a new estimate of denominator $d^{(i)}(s)$, and
	\item compensating for the approximation introduced by fixing the denominator of $\widetilde{H}^{(i)}(s)$ to the previous iteration value.  Indeed, weight $w^{(i)}(s)$ depends on the ratio between new denominator estimate $d^{(i)}(s)$ and previous estimate $d^{(i-1)}(s)$.
\end{itemize} 

Next, we derive alternative expressions for  $w^{(i)}(s)$ and $\widetilde{H}^{(i)}(s)$, which pave the way for an implicit weighting of~\eqref{eq:vf_eSK2}. Substituting~\eqref{eq:vf_n} and~\eqref{eq:vf_d} into~\eqref{eq:vf_wbefore} we can derive the following chain of equalities
\begin{equation}
	w^{(i)}(s) 
		= \frac{1 + \sum_{n=1}^{\bar{n}} \frac{d_n^{(i)}}{s-p_n^{(0)}}}{1 + \sum_{n=1}^{\bar{n}} \frac{d_n^{(i-1)}}{s-p_n^{(0)}}} 
		= \frac{\frac{\prod \left(s-p_n^{(i)}\right)}{\prod \left(s-p_n^{(0)}\right)}}{\frac{\prod \left(s-p_n^{(i-1)}\right)}{\prod \left(s-p_n^{(0)}\right)}}
		= 1+  \sum_{n=1}^{\bar{n}} \frac{w_n^{(i)}}{s-p_n^{(i-1)}}
		\,,
	\label{eq:vf_w}
\end{equation}
where $\prod = \prod_{n=1}^{\bar{n}}$. In~\eqref{eq:vf_w}, $p_n^{(i)}$ are the \textit{zeros} of $d^{(i)}(s)$, and therefore the \emph{poles} of $\widetilde{H}^{(i)}(s)$. From the second expression in~\eqref{eq:vf_w}, we see that $w^{(i)}(s)$ is the ratio of two rational functions with the same poles $p_n^{(0)}$. By factorizing their respective numerators and denominators, as in the third expression, we observe that those common poles can be eliminated. Finally, we express $w^{(i)}$ in terms of a new set of poles $p_n^{(i-1)}$ that change at every iteration, as in the last expression in~\eqref{eq:vf_w}. The same manipulation can be performed on $\widetilde{H}^{(i)}(s)$, leading to
\begin{equation}
	\widetilde{H}^{(i)}(s) = \frac{c_0^{(i)} + \sum_{n=1}^{\bar{n}} \frac{c_n^{(i)}}{s-p_n^{(0)}}}{1 + \sum_{n=1}^{\bar{n}} \frac{d_n^{(i-1)}}{s-p_n^{(0)}}} 
	= r_0^{(i)} +  \sum_{n=1}^{\bar{n}} \frac{r_n^{(i)}}{s-p_n^{(i-1)}}\,.
	\label{eq:vf_H}
\end{equation}
Substituting~\eqref{eq:vf_w} and~\eqref{eq:vf_H} into~\eqref{eq:vf_eSK2}, we obtain~\cite{gustavsen1999rational,grivet2015passive}
\begin{equation}
	\left( e_{SK}^{(i)} \right)^2 = \frac{1}{\bar{k}}  \sum_{k=1}^{\bar{k}} \left| 
			H_k \left( 1+  \sum_{n=1}^{\bar{n}} \frac{w_n^{(i)}}{\jmath \omega_k-p_n^{(i-1)}} \right) - 
			\left( r_0^{(i)} +  \sum_{n=1}^{\bar{n}} \frac{r_n^{(i)}}{\jmath \omega_k-p_n^{(i-1)}} \right)
			\right|^2\,,
	\label{eq:vf_eVF}
\end{equation}
which is the actual error function used in VF to fit the model to the given samples. The main difference between~\eqref{eq:vf_eVF} and~\eqref{eq:vf_eSK} is how the linearized error is iteratively weighted to progressively converge to~\eqref{eq:vf_e}. In~\eqref{eq:vf_eSK2}, weight $\frac{1}{d^{(i-1)}(\jmath \omega_k)}$ is applied \emph{explicitly}, which degrades numerical conditioning. In~\eqref{eq:vf_eVF}, instead, the same weight is applied \emph{implicitly} by relocating the poles $p_n^{(i-1)}$ at each iteration. 

Once~\eqref{eq:vf_eVF} has been minimized, the updated poles $p_n^{(i)}$ for the next iteration can be found as the zeros of $d^{(i)}(s)$, as one can see from the third expression in~\eqref{eq:vf_w}. It can be shown that such zeros can be calculated as the eigenvalues of~\cite{gustavsen1999rational}
\begin{equation}
	\left\{ p_n^{(i)} \right\} = \text{eig} \left( A^{(i-1)}  - b_w \left(c^{(i)}_w \right)^T \right) \,,
	\label{eq:vf_matrix4poles}
\end{equation}
with $A^{(i-1)} = \text{diag} \left \{ p_1^{(i-1)},\dots,p_{\bar{n}}^{(i-1)} \right \}$ being a diagonal matrix formed by poles $p_n^{(i-1)}$. In~\eqref{eq:vf_matrix4poles} $b_w$ is a $\bar{n} \times 1$ vector of ones, and $\left(c^{(i)}_w \right)^T = [w_1^{(i)},\dots,w_{\bar{n}}^{(i)}]$. Upon convergence, $p_n^{(i-1)} \to p_n^{(i)}$, and become the poles of the obtained model $H^{(i)}(s)$. When this happens, $w^{(i)} \to 1$ as we can see from the third expression in~\eqref{eq:vf_w}, and linearized error~\eqref{eq:vf_eSK2} tends to~\eqref{eq:vf_e}, as desired.

\subsection{The Vector Fitting algorithm}
\label{sec:vf_siso}

We are now ready to present the complete VF algorithm~\cite{gustavsen1999rational,grivet2015passive}, with a pseudo-code implementation available in Algorithm~\ref{alg:vf_vf}. The first step is to choose the order $\bar{n}$ of the desired model. This choice will be discussed in Sec.~\ref{sec:vf_orderestimation}. Next, we set the initial poles $p_n^{(0)}$ of the basis functions in~\eqref{eq:vf_H} and~\eqref{eq:vf_w}. Numerical tests~\cite{gustavsen1999rational} showed that a linear distribution of poles with small and negative real part over the bandwidth spanned by samples $H_k$ leads to the best conditioning of the least squares problems to be solved. We assume $\bar{n}$ even, and frequency values $\omega_k$ sorted in ascending order. If $\omega_1=0$, the initial poles can be set as~\cite{grivet2015passive}
 \begin{equation}
	 p_n^{(0)} = 
	 \begin{cases}
		 (-\alpha + \jmath) \frac{\omega_{\bar{k}}}{\bar{n}/2} n  & \text{ for } n=1,\dots, \bar{n}/2  \\
		 \left( p_{n-\bar{n}/2}^{(0)} \right)^* & \text{ for } n = \bar{n}/2 + 1, \dots, \bar{n}
	 \end{cases}				
 	\label{eq:vf_p0DC}
 \end{equation}
where $^*$ denotes the complex conjugate and $\alpha$ is typically set to 0.01. This rule generates $\bar{n}/2$ pairs of complex conjugate poles, linearly distributed over the frequency range $[0,\omega_{\bar{k}}]$ spanned by samples $H_k$. The imaginary part of the poles is set to be quite larger than the real part, since this makes the partial fraction basis functions well distinct from each other, which improves numerical conditioning. 

When $\omega_1 \neq 0$, distribution~\eqref{eq:vf_p0DC} can be modified as~\cite{grivet2015passive}
\begin{equation}
p_n^{(0)} = 
\begin{cases}
(-\alpha + \jmath) \left[ \omega_1 + \frac{\omega_{\bar{k}} - \omega_1}{\bar{n}/2-1} (n-1)  \right] & \text{ for } n=1,\dots, \bar{n}/2  \\
\left( p_{n-\bar{n}/2}^{(0)} \right)^* & \text{ for } n = \bar{n}/2 + 1, \dots, \bar{n}
\end{cases}				
\label{eq:vf_p0}
\end{equation}
to linearly spread the poles between $\omega  = \omega_1$ and $\omega = \omega_{\bar{k}}$. Rules~\eqref{eq:vf_p0DC} and~\eqref{eq:vf_p0} work well for most cases, since the choice of initial poles is typically not critical for VF convergence. When the frequency range of interest spans several decades, and the system frequency response exhibits significant behavior in multiple decades, initial poles can be distributed logarithmically for optimal results~\cite{grivet2015passive}.

The core of the VF algorithm is an iterative minimization of~\eqref{eq:vf_eVF}, which begins with $i=1$. Minimizing~\eqref{eq:vf_eVF} is equivalent to solving, in least-squares sense, the system of equations
\begin{equation}
	\begin{bmatrix}
		\Phi_0^{(i)} & -D_H \Phi_1^{(i)}
	\end{bmatrix}
	\begin{bmatrix}
		 c^{(i)}_H \\ c^{(i)}_w
	\end{bmatrix}	
	= V_H
	\label{eq:vf_sys}
\end{equation}
where $\Phi_0^{(i)}$ and $\Phi_1^{(i)}$ contain the partial fraction basis functions evaluated at the different frequency points $\omega_k$
\begin{align}
	\Phi_0^{(i)} & = 
	\begin{bmatrix}
		1	& \frac{1}{\jmath \omega_1 - p_1^{(i-1)}} & \dots & \frac{1}{\jmath \omega_1 - p_{\bar{n}}^{(i-1)}}  \\
		\vdots & \vdots & & \vdots \\
		1	& \frac{1}{\jmath \omega_{\bar{k}} - p_1^{(i-1)}} & \dots & \frac{1}{\jmath \omega_{\bar{k}} - p_{\bar{n}}^{(i-1)}} \\
	\end{bmatrix}\,,
	\label{eq:vf_Phi0}\\
	\Phi_1^{(i)} & = 
	\begin{bmatrix}
		\frac{1}{\jmath \omega_1 - p_1^{(i-1)}} & \dots & \frac{1}{\jmath \omega_1 - p_{\bar{n}}^{(i-1)}}\\
		\vdots & & \vdots\\
		\frac{1}{\jmath \omega_{\bar{k}} - p_1^{(i-1)}} & \dots & \frac{1}{\jmath \omega_{\bar{k}} - p_{\bar{n}}^{(i-1)}}\\
	\end{bmatrix}\,,
	\label{eq:vf_Phi1}
	\end{align}
and $D_H = \text{diag} \{H_1, \dots, H_{\bar{k}}\}$. The right hand side of~\eqref{eq:vf_sys} is a column vector formed by the given samples
\begin{equation}
		V_H = 
	\begin{bmatrix}
	H_1 & \dots & H_{\bar{k}}
	\end{bmatrix}^T\,,
	\label{eq:vf_rhs}
\end{equation}
and $c_H^{(i)}$ and $c_w^{(i)}$ contain the unknown coefficients
\begin{align}
	c_H^{(i)} & = \begin{bmatrix}
														r_0^{(i)} & \dots & r_{\bar{n}}^{(i)} 
													\end{bmatrix}^T\,,
	\label{eq:vf_cH} \\
	c_w^{(i)} & = 
	\begin{bmatrix}
		w_1^{(i)} & \dots & w_{\bar{n}}^{(i)} 
	\end{bmatrix}^T\,.
	\label{eq:vf_cw}
\end{align}
System~\eqref{eq:vf_sys} can be solved in least-squares sense with a QR decomposition of the coefficient matrix~\cite{Gol96}. Once~\eqref{eq:vf_sys} has been solved, the new poles estimate $p_n^{(i)}$ is computed with~\eqref{eq:vf_matrix4poles}.

The VF iterative process usually converges very quickly, often in 4-5 iterations, except when the given samples are noisy. The fast and reliable convergence of VF is truly remarkable considering that VF ultimately solves a nonlinear minimization problem. Unfortunately, so far no one has been able to support this experimental evidence with strong theoretical results on VF convergence. Actually, contrived examples show that VF convergence is not guaranteed~\cite{lefteriu2013convergence,shi2016nonconvergence}. However, these examples are quite artificial and far from practical datasets. Two decades of widespread use indeed show that, when properly implemented, VF is a remarkably robust algorithm for the identification of reduced-order models from sampled data. In VF, convergence is typically monitored with three conditions:
\begin{enumerate}
		\item when pole estimates stabilize, i.e. $p_n^{(i)} \simeq p_n^{(i-1)}$, performing new iterations will not improve accuracy. When this happens, $w^{(i)}(\jmath \omega) \simeq 1$ for $\omega \in [\omega_1, \omega_{\bar{k}}]$. This occurrence can be tested numerically as
	\begin{equation}
			\max_k \vfabs{w^{(i)}(\jmath \omega_k)-1} \le \varepsilon_w			 \,,
		\label{eq:vf_convw}
	\end{equation} 
	where $\varepsilon_w$ is a user-defined threshold. The advantage of criterion~\eqref{eq:vf_convw} is that it does not require additional computations apart from the calculation of the norm of $w'$.  The limitation is that this condition only checks if the iterative process has stabilized, which does not necessarily mean that $H^{(i)}(s)$ fits well the given frequency samples;
	\item when condition~\eqref{eq:vf_convw} is satisfied, the error between the fitted model and samples $H_k$ should be checked. In principle, this can be done by computing the error between~\eqref{eq:vf_H} and $H_k$. However, since after solving~\eqref{eq:vf_sys} a new estimate of the poles can be found via~\eqref{eq:vf_matrix4poles}, the common practice is to use those poles to fit a new model. This is done by minimizing the exact error~\eqref{eq:vf_e} between the given samples $H_k$ and model
	\begin{equation}
		\widetilde{H}^{(i+1)}(s)  =  r_0^{(i+1)} +  \sum_{n=1}^{\bar{n}} \frac{r_n^{(i+1)}}{s-p_n^{(i)}}\,,
		\label{eq:vf_Hvf}
	\end{equation}
	considering only residues $r_0^{(i+1)},\dots, r_{\bar{n}}^{(i+1)}$ as unknowns. Since poles $p_n^{(i)}$ are now fixed, this is equivalent to solve, in least squares sense, the linear system
	\begin{equation}
		\Phi_0^{(i+1)} c_H^{(i+1)} = V_H\,.
		\label{eq:vf_sysfixedpoles}
	\end{equation}
	 The VF iteration ends, successfully, when
		\begin{equation}
			e \le \varepsilon_H\,,
			\label{eq:vf_convH}
		\end{equation}
		since model~\eqref{eq:vf_Hvf} meets the accuracy threshold $\varepsilon_H$ set by the user. 
		The main reason why this additional fitting step is performed is because this step minimizes the exact error~\eqref{eq:vf_e} between model and given samples, rather then a linear approximation like~\eqref{eq:vf_eVF}, which improves accuracy and more reliably detects convergence. Therefore, solving~\eqref{eq:vf_sysfixedpoles} serves both as convergence test and as final fitting of the model;
	\item in selected circumstances, VF may be unable to reach~\eqref{eq:vf_convH} even after many iterations. In this case, the iterative process concludes unsuccessfully when $i$ exceeds the maximum number of iterations $i_{max}$ allowed by the user.
\end{enumerate}

\begin{algorithm}[t]
	\caption{Vector Fitting}
	\label{alg:vf_vf}
	\begin{algorithmic}[1]
		\Require response samples $H_k$, corresponding frequencies $\omega_k$ ($k=1,\dots,\bar{k})$
		\Require desired model order $\bar{n}$
		\Require maximum number of iterations $i_{max}$
		\State set initial poles $p_n^{(0)}$ according to~\eqref{eq:vf_p0DC} or~\eqref{eq:vf_p0}.
		\State $i \gets 1$
		\While{$i \le i_{max}$}
			\State Solve~\eqref{eq:vf_sys} or~\eqref{eq:vf_sysmimo} in least squares sense \label{alg:vf_fit1}
			\State Compute the new poles estimate $p_n^{(i)}$ with~\eqref{eq:vf_matrix4poles} \label{alg:vf_np}
			\State Enforce poles stability with~\eqref{eq:vf_stablepoles}, if desired \Comment Stability enforcement
			\If{\eqref{eq:vf_convw} is true} \Comment First convergence test
				\State Solve~\eqref{eq:vf_sysfixedpoles} or~\eqref{eq:vf_sysfixedpolesmimo} in least squares sense \label{alg:vf_fit2} \Comment Tentative final fitting
				\State Compute fitting error $e$ with~\eqref{eq:vf_e} or~\eqref{eq:vf_emimo}
				\If{$e \le \varepsilon_H$} \Comment Second convergence test
					\State $\widetilde{H}(s) = \widetilde{H}^{(i+1)}(s)$
					\State \Return Success!
				\EndIf
			\EndIf
			\State $i \gets i+1$
		\EndWhile
		\State \Return Failure: maximum number of iterations reached.
	\end{algorithmic}
\end{algorithm}
\clearpage

\subsection{Example: fitting a rational transfer function}
\label{sec:vf_ex1}

\begin{table}[t]
	\centering
	\begin{tabular}{c|c}
		
		Pole & Residue \\ 
		\hline 
		constant term & $r_0 = 0.1059$ \\ 
		
		$p_1 = -1.3578$ & $r_1 = -0.2808$ \\ 
		
		$p_2 = -1.2679$ & $r_2 = 0.1166$ \\ 
		
		$p_{3,4} = -1.4851 \pm 0.2443\jmath$ & $r_{3,4} = 0.9569 \mp 0.7639\jmath$ \\ 
		
		$p_{5,6} = -0.8487 \pm 2.9019\jmath$ & $r_{5,6} = 0.9357 \mp 0.7593\jmath$ \\ 
		
		$p_{7,8} = -0.8587 \pm 3.1752\jmath$ & $r_{7,8} = 0.4579 \mp 0.7406\jmath$ \\ 
		
		$p_{9,10} = -0.2497 \pm 6.5369\jmath$ & $r_{9,10} = 0.2405 \mp 0.7437\jmath$ \\ 
	\end{tabular} 
	\caption{Example of Sec.~\ref{sec:vf_ex1}: poles and residues of the transfer function used to generate samples $H_k$.}
	\label{tab:vf_ex1}
\end{table}

\begin{figure}
	\includegraphics[width=0.45\linewidth]{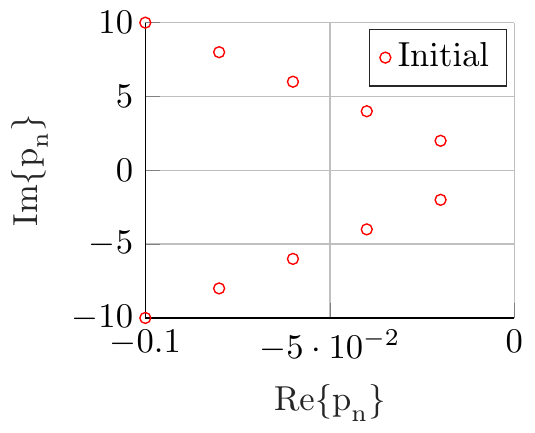}
	\hfill
	\includegraphics[width=0.44\linewidth]{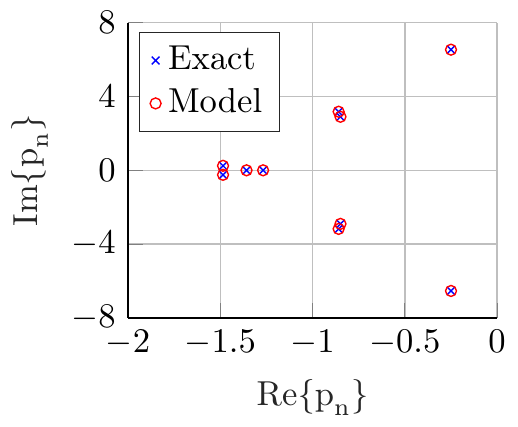}
	\caption{Left panel: initial poles $p_n^{0}$ used by VF in the first iteration. Right panel: poles of the final model $\widetilde{H}(s)$ compared to the exact poles of the original transfer function.}
	\label{fig:vf_ex1_poles}
\end{figure}

In this example, we apply VF to a set of samples $H_k$ generated from a known rational function of order 10. Its poles were generated randomly, and  are reported  in Table~\ref{tab:vf_ex1}. The original transfer function was sampled at $\bar{k}=100$ frequency points linearly spaced between $\omega_1 = 0.1\,{\rm rad/s}$ and $\omega_{100} = 10\,{\rm rad/s}$. A Matlab implementation of VF was used to fit the samples with a model in the form~\eqref{eq:vf_H} with order $\bar{n} = 10$. The initial distribution of poles $p_n^{(0)}$ set by~\eqref{eq:vf_p0} is depicted in the left panel of Fig.~\ref{fig:vf_ex1_poles}. Throughout the VF iterations, poles relocate to the final distribution shown in the right panel of Fig.~\ref{fig:vf_ex1_poles}, which also compares them to the exact poles of the original rational function. We can see that the poles estimated by VF closely match the poles of the original system. 

\begin{figure}
	\begin{center}
		~\hspace{0.33cm}~\includegraphics[width=0.76\linewidth]{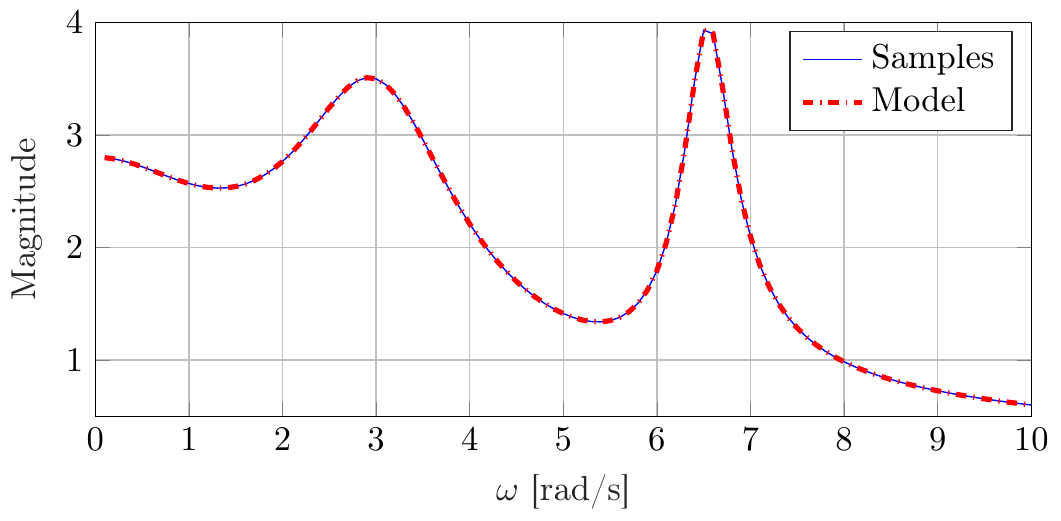}
	\end{center}
	\begin{center}
			\includegraphics[width=0.8\linewidth]{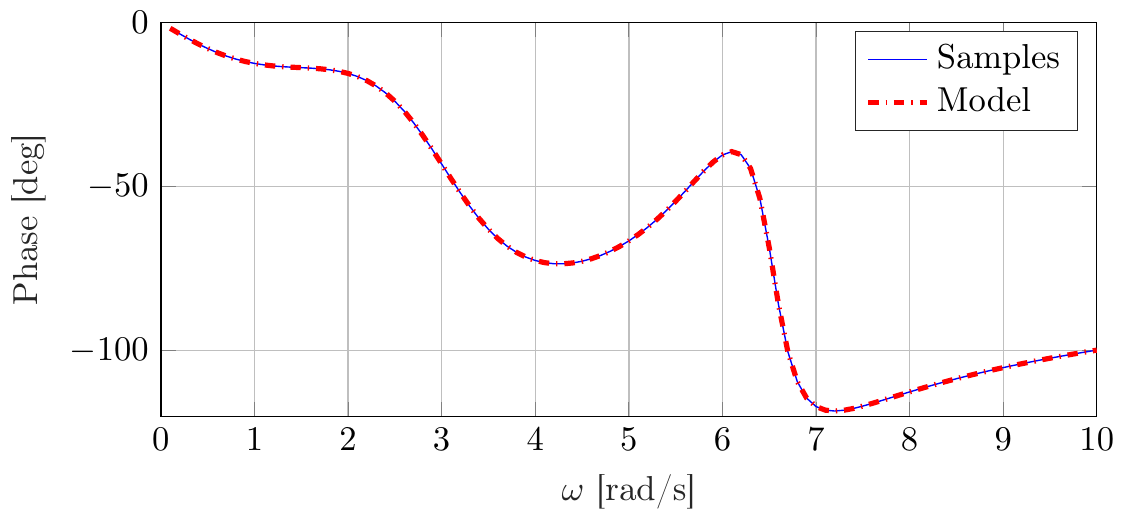}~
	\end{center}
	\caption{Example of Sec.~\ref{sec:vf_ex1}: magnitude (top) and phase (bottom) of samples $H_k$ and of the model $\widetilde{H}(\jmath \omega)$ identified by VF.}
	\label{fig:vf_ex1_h}
\end{figure}

\begin{figure}
	\centering
	\includegraphics[width=0.8\linewidth]{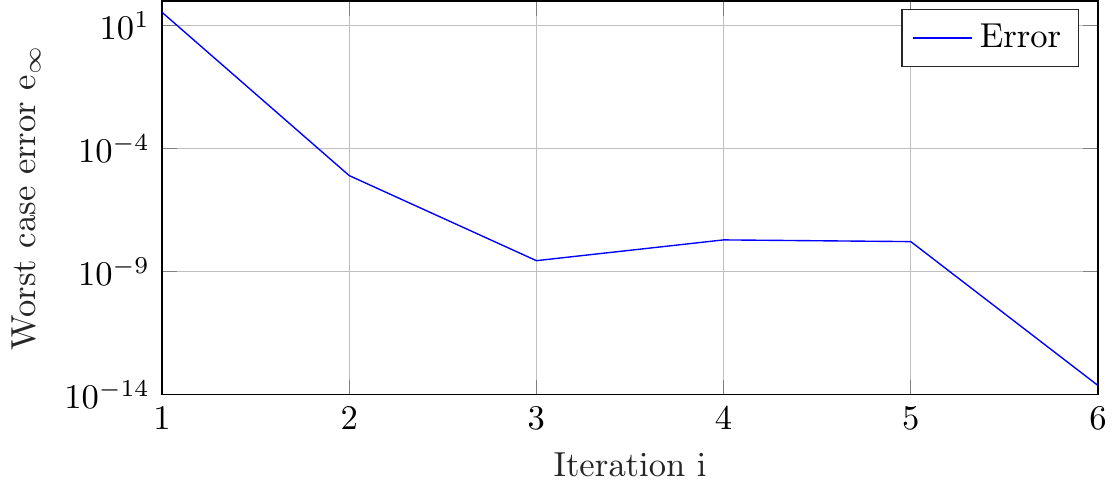}
	\caption{Example of Sec.~\ref{sec:vf_ex1}: worst-case fitting error $e_{\infty}$ as a function of iteration counter $i$. The last iteration ($i=6$) was performed with fixed poles.}
	\label{fig:vf_ex1_conv}
\end{figure}

In Figure~\ref{fig:vf_ex1_h}, the frequency response $\widetilde{H}(\jmath \omega)$ of the VF model is compared to the initial samples. We observe an excellent agreement over the entire frequency range of interest. At the conclusion of the VF iterative process, the worst case error between samples $H_k$ and model response
\begin{equation}
	e_{\infty} = \max_k \left| H_k - H(\jmath \omega_k)  \right|
	\label{eq:vf_eWC}
\end{equation}
is $2.37 \times 10^{-14}$. Figure~\ref{fig:vf_ex1_conv} shows the evolution of $e_{\infty}$ throughout the five iterations performed by VF, plus a final  iteration ($i=6$) where poles were kept fixed and residues were calculated one more time using~\eqref{eq:vf_sysfixedpoles}. The Figure shows that VF converges very quickly, reaching an error below $10^{-8}$ in only three iterations. We can also observe that the final fitting iteration ($i=6$) with fixed poles provides a more accurate model. For this example, VF took only 0.2~s of CPU time on a 2.2~GHz mobile processor. The source codes related to this example can be downloaded from~\cite{myvfwebsite}.

\subsection{Example: modeling of aortic input impedance}
\label{sec:vf_ex2}

In this example, VF is used to model the relation between pressure $p(t)$ and flow rate $q(t)$ in the ascending aorta of a 1.1-year old patient~\cite[patient 1]{sharp2000aortic}. Simultaneous pressure and flow rate measurements were collected during a surgical procedure. Blood flow rate was measured with an ultrasonic flow probe positioned about 1~cm downstream of the aortic valve. Pressure was acquired using a catheter with a pressure transducer on its tip, positioned in the same location as the flow rate probe. From time-domain recordings, the input impedance seen from the aorta was obtained as 
\begin{equation}
	Z(\jmath \omega) = \frac{\mathscr{F}\{p(t)\}}{\mathscr{F}\{q(t)\}}\,,
	\label{eq:vf_Z}
\end{equation}
where $\mathscr{F}\{.\}$ denotes the Fourier transform. Impedance was computed at $\bar{k}=11$ frequency points $\omega_k = 2 \pi (k-1) f_0$ for $k=1,...,11$, where $f_0 = 2.54\,{\rm Hz} = 152.4 \,{\rm beats/min}$ corresponds to the heart rate of the patient. The authors of~\cite{sharp2000aortic} estimate that the impedance measurements are affected by uncertainty with a relative standard deviation that ranges between 0.66\% to 14.5\% depending on frequency. Relative standard deviation was normalized to $\left| Z(0)\right|$.

\begin{figure}
	\includegraphics[width=0.52\linewidth]{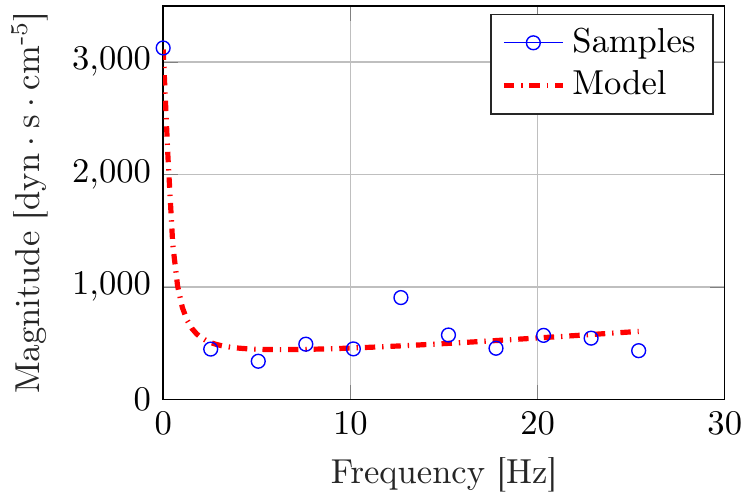}
	\hfill
	\includegraphics[width=0.5\linewidth]{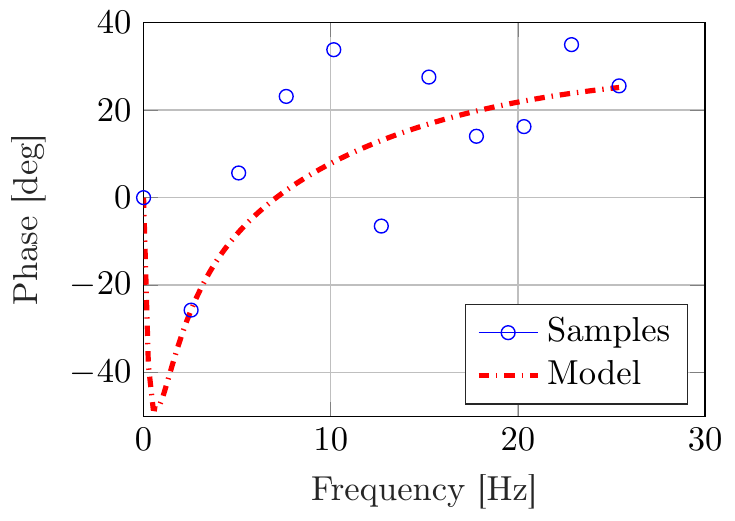}
	
	\includegraphics[width=0.52\linewidth]{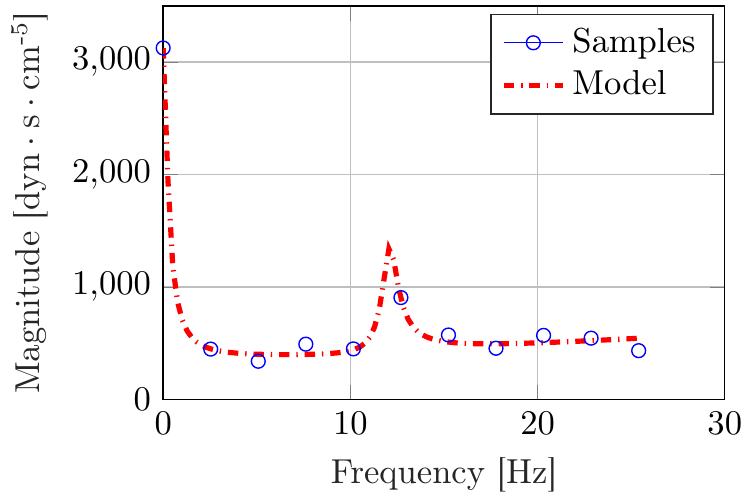}
	\hfill
	\includegraphics[width=0.5\linewidth]{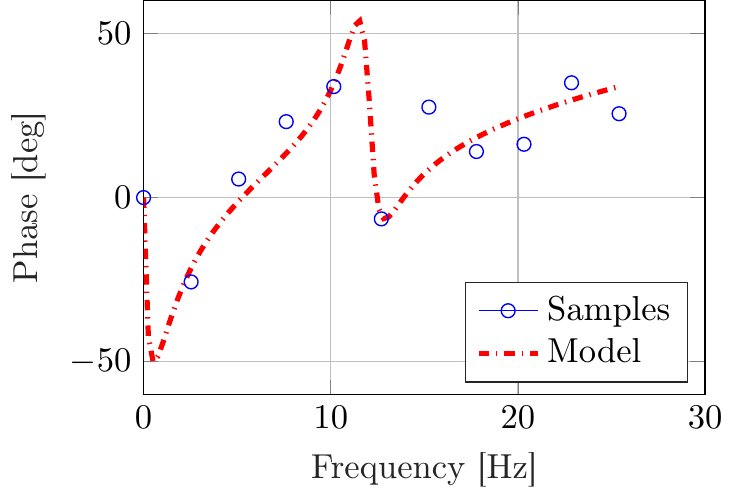}
	
	\includegraphics[width=0.52\linewidth]{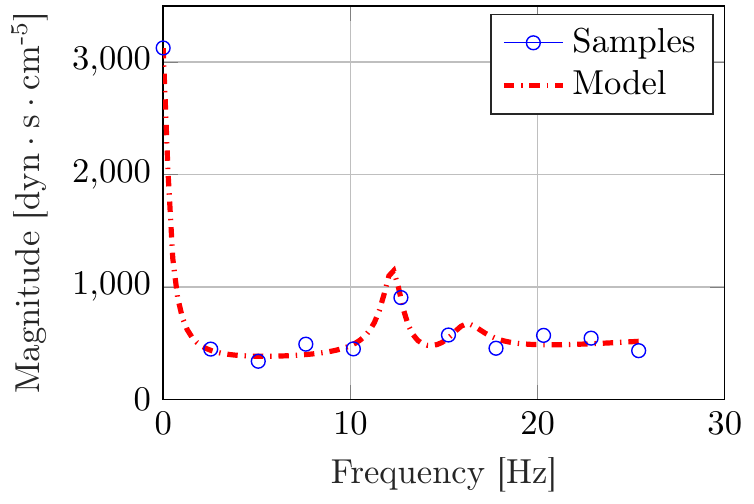}
	\hfill
	\includegraphics[width=0.5\linewidth]{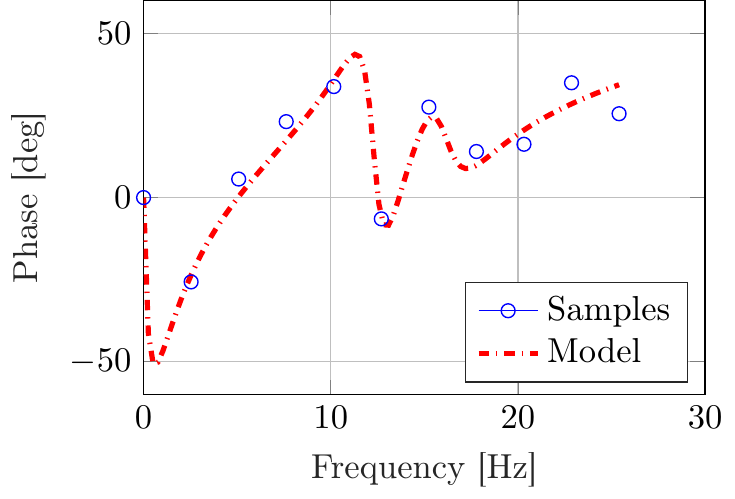}
	
	\includegraphics[width=0.52\linewidth]{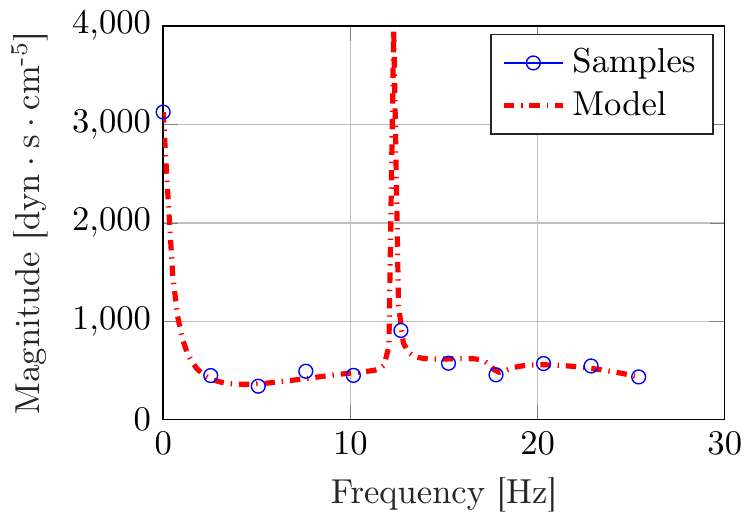}
	\hfill
	\includegraphics[width=0.5\linewidth]{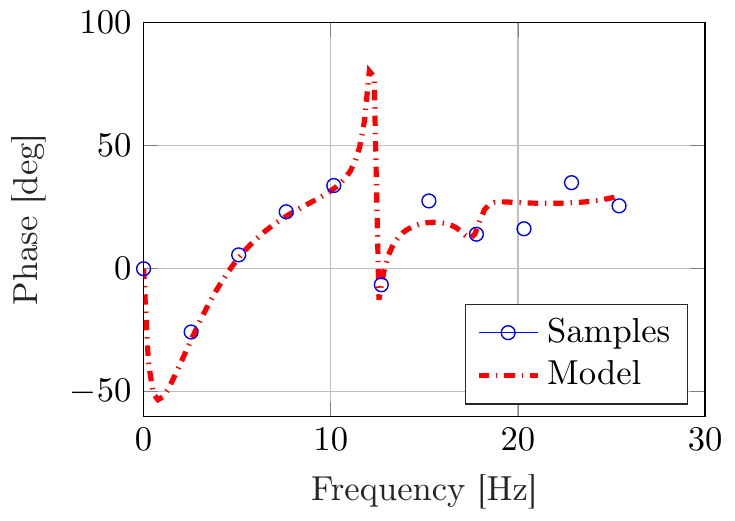}
	\caption{Impedance seen into the ascending aorta of the pediatric patient considered in Sec.~\ref{sec:vf_ex2}: measured samples (circles) and response of four different VF models (dashed lines) of order $\bar{n} = 2, 4, 6, 8$ (from top to bottom).}
	\label{fig:vf_ex2}
\end{figure}

We apply VF to the impedance samples to obtain a closed-form model relating aortic pressure and flow rate. The limited number of available samples, and their uncertainty, make the identification of an accurate model challenging. We use this non-trivial scenario to explore the relation between number and quality of the available samples, model order $\bar{n}$, and accuracy. Vector Fitting was applied to the given samples four times with model order $\bar{n}$ increasing from 2 to 8 in steps of 2. Figure~\ref{fig:vf_ex2} compares the magnitude and phase of the identified model to the original impedance samples. We can see that the $\bar{n} = 2$ model captures the overall trend of the impedance. However, it fails to resolve the increase in impedance at $f=12.7 \,{\rm Hz}$ and the associated phase variation. Increasing order to 4 or 6  resolves that feature and provides higher accuracy. Further increasing order $\bar{n}$ to 8 leads to a model which matches closely most given samples, but has a sharp and high peak at $f=12.3\,{\rm Hz}$. This unrealistic behavior in-between the given samples is typical of an overfitting scenario, where the sought model has too many degrees of freedom, which can be hardly estimated from the information contained in the available samples. Although still solvable, the conditioning number of~\eqref{eq:vf_sys} degrades. The system solution, which gives the model coefficients, becomes very sensitive to the noise superimposed to the given samples. The source codes related to this example can be downloaded from~\cite{myvfwebsite}.

\subsection{The multi-input multi-output case}
\label{sec:vf_mimo}

The VF algorithm presented in Sec.~\ref{sec:vf_siso} for the single-input single-output case can be easily extended to the general case of a system with $\bar{m}$ inputs and $\bar{q}$ outputs. In this case, the given samples are $\bar{q} \times \bar{m}$ complex matrices $H_k$, and we denote their $(q,m)$ entry as $H_{k,qm}$. The model transfer function is now defined as
\begin{equation}
	\widetilde{H}^{(i)}(s) 	= R_0^{(i)} +  \sum_{n=1}^{\bar{n}} \frac{R_n^{(i)}}{s-p_n^{(i-1)}}\,.
	\label{eq:vf_Hmimo}
\end{equation}
where $R_n^{(i)} \in \mathbb{C}^{\bar{q} \times \bar{m}}$. In~\eqref{eq:vf_Hmimo}, the same poles $p_n^{(i-1)}$ are used for all elements of matrix $\widetilde{H}^{(i)}(s)$. This choice is appropriate when modeling linear dynamical systems, since it is known that the poles of each transfer function entry are a subset of a common set of poles shared by all transfer function elements. The physical justification of this fact is that poles are related to the natural modes of the system, which are a property of the system itself and not of individual entries of its transfer function. In other communities, natural modes are referred to as resonances or eigenmodes of the system. When VF is applied to model transfer functions not related to the same physical system, one should use distinct poles for different elements of~\eqref{eq:vf_Hmimo}. This scenario is discussed in~\cite{grivet2015passive}, which also elaborates on the computational implications of this choice.

In the multi-input multi-output case,  weighting function $w^{(i)}(s)$ remains defined by~\eqref{eq:vf_w}. Since transfer function~\eqref{eq:vf_Hmimo} is now matrix-valued, VF aims to minimize the error functional
\begin{equation}
	e^2 = \frac{1}{\bar{k} \bar{q} \bar{m}} \sum_{k=1}^{\bar{k}} \vfnorm{H_k - \widetilde{H}(\jmath \omega_k)}^2_F\,,
	\label{eq:vf_emimo}
\end{equation}
where $\vfnorm{.}_F$ denotes the Frobenius norm, which for $A \in \mathbb{C}^{\bar{q} \times \bar{m}}$ is defined as
\begin{equation}
	\vfnorm{A}_F = \sqrt{\sum_{q=1}^{\bar{q}} \sum_{m=1}^{\bar{m}}  \left| A_{qm} \right|^2}\,.
	\label{eq:frobenius}
\end{equation}
From~\eqref{eq:frobenius}, we see that the square of the Frobenius norm is simply equal to the sum of the squared magnitude of each entry. Therefore, minimizing~\eqref{eq:vf_emimo} means minimizing the sum of the squared error between each sample $H_{k,qm}$ and the corresponding entry of~\eqref{eq:vf_Hmimo}. 

The minimization of~\eqref{eq:vf_emimo} is a nonlinear least squares problem, which VF solves iteratively by working on the linearized error~\cite{gustavsen1999rational}
\begin{equation}
	\left( e_{SK}^{(i)} \right)^2 = \frac{1}{\bar{k} \bar{q} \bar{m}} \sum_{k=1}^{\bar{k}} \vfnorm{
	H_k \left( 1+  \sum_{n=1}^{\bar{n}} \frac{w_n^{(i)}}{\jmath \omega_k-p_n^{(i-1)}} \right) - 
	\left( R_0^{(i)} +  \sum_{n=1}^{\bar{n}} \frac{R_n^{(i)}}{\jmath \omega_k-p_n^{(i-1)}} \right)
	}_F^2\,.
	\label{eq:vf_eVFmimo}
\end{equation}
As in the single-input single-output case, we can see that~\eqref{eq:vf_eVFmimo} uses weighting function $w^{(i)}(s)$ to offset the error introduced by using the previous poles estimate in the denominators. Minimizing~\eqref{eq:vf_eVFmimo} is equivalent to solving, in least squares sense, the system of equations
\begin{equation}
	R_{0,qm}^{(i)} + \sum_{n=1}^{\bar{n}} \frac{R_{n,qm}^{(i)}}{\jmath \omega_k - p_n^{(i-1)}}
	- H_{k,qm} \sum_{n=1}^{\bar{n}} \frac{w_n^{(i)}}{\jmath \omega_k-p_n^{(i-1)}}
	= H_{k,qm}
	\label{eq:vf_eqmimo}
\end{equation}
for $k =1, \dots, \bar{k}$, $q = 1,\dots, \bar{q}$ and $m = 1,\dots, \bar{m}$. In matrix form, equations~\eqref{eq:vf_eqmimo} read
\begin{equation}
	\begin{bmatrix}
		\Phi_0^{(i)}	&	0 & \dots & 0  &-D_{H_{11}} \Phi_1^{(i)} \\
		0					&		\Phi_0^{(i)} & \ddots & \vdots & -D_{H_{21}} \Phi_1^{(i)} \\
		\vdots & \ddots & \ddots & 0 & \vdots\\
		0 & \dots & 0 & \Phi_0^{(i)} & -D_{H_{\bar{q} \bar{m}}} \Phi_1^{(i)} \\
	\end{bmatrix}
	\begin{bmatrix}
		c_{H_{11}}^{(i)} \\		
		c_{H_{21}}^{(i)} \\
		\vdots \\
				c_{H_{\bar{q} \bar{m}}}^{(i)} \\
				c_w^{(i)}\\
	\end{bmatrix}
	=
	\begin{bmatrix}
		V_{H_{11}} \\
		V_{H_{21}} \\
		\vdots \\
		V_{H_{\bar{q} \bar{m}}}
	\end{bmatrix}\,,
	\label{eq:vf_sysmimo}
\end{equation}
where $D_{H_{qm}}$ and $V_{H_{qm}}$ are, respectively, a diagonal matrix and a column vector formed by all samples $H_{k,qm}$ for $k=1,\dots,\bar{k}$. In the unknown vector of~\eqref{eq:vf_sysmimo},
\begin{equation}
	c_{H_{qm}}^{(i)} = 
	\begin{bmatrix}
		R_{0,qm}^{(i)} & \dots & R_{\bar{n},qm}^{(i)} 
	\end{bmatrix}^T\,,
	\label{eq:vf_cHqm}
\end{equation}
and $c_w^{(i)}$ is defined by~\eqref{eq:vf_cw}. System~\eqref{eq:vf_sysmimo} is solved in step~\ref{alg:vf_fit1} of Algorithm~\ref{alg:vf_vf}. In step~\ref{alg:vf_fit2}, a tentative final fitting of the model is performed, assuming fixed poles and determining only a new estimate of residues $R_{n}^{(i+1)}$. This step can be achieved by solving
\begin{equation}
	\Phi_0^{(i+1)} c_{H_{qm}}^{(i+1)} = V_{H_{qm}}\,,
	\label{eq:vf_sysfixedpolesmimo}
\end{equation} 
for $q=1,\dots,\bar{q}$ and $m=1,\dots,\bar{m}$.

\subsection{The fast Vector Fitting algorithm}
\label{sec:vf_fastvf}

As the number of inputs $\bar{m}$ and outputs $\bar{q}$ increases, the computational cost of solving~\eqref{eq:vf_sysmimo} can quickly become unsustainable. As technology evolves, this scenario arises more frequently, as engineers need to model systems of increasing complexity, either in terms of dynamic order or number of inputs and outputs. For example, a modern server processor has about 2,000 pins, which are connected to the motherboard by a dense network of tiny wires realized on the chip package. Seen as an input-output system, this network will have about 4,000 inputs and outputs, half where the network connects to the motherboard, and half where the network is connected to the silicon die. The need to predict electromagnetic interference in this dense and intricate network of wires calls for scalable algorithms to create reduced-order models for systems where the number of inputs $\bar{m}$ and outputs $\bar{q}$ can be several thousands~\cite{schilders2011need,baur2014model}. 

 The Fast VF algorithm~\cite{deschrijver2008macromodeling,knockaert2009comments} significantly reduces the cost of solving~\eqref{eq:vf_sysmimo}  for multi-input and multi-output systems. Savings are achieved by exploiting the block structure of~\eqref{eq:vf_sysmimo} and the fact that, of the solution vector of~\eqref{eq:vf_sysmimo}, only $c_w^{(i)}$ is actually needed to compute the new poles estimate~\eqref{eq:vf_matrix4poles}. Least squares problem in the form~\eqref{eq:vf_sysmimo} can be efficiently solved by first performing the QR decompositions~\cite{golub1980large, bayard1994high,verboven2005multivariable,deschrijver2008macromodeling}
\begin{equation}
\begin{bmatrix}
	\Phi_0^{(i)}  & -D_{H_{qm}} \Phi_1^{(i)} \\
\end{bmatrix}
= 
\begin{bmatrix}
	{\cal Q}_{qm}^{1} & {\cal Q}_{qm}^{2}
\end{bmatrix}
\begin{bmatrix}
	{\cal R}_{qm}^{11}	& 	{\cal R}_{qm}^{12} \\
	0						& 	{\cal R}_{qm}^{22}
\end{bmatrix}
\,,
\label{eq:vf_fastvf1}
\end{equation}
for $q=1,\dots,\bar{q}$ and $m=1,\dots,\bar{m}$. Then, a reduced system is formed~\cite{deschrijver2008macromodeling}
\begin{equation}
\begin{bmatrix}
	{\cal R}_{11}^{22} \\
	{\cal R}_{21}^{22} \\
	\vdots \\
	{\cal R}_{\bar{q}\bar{m}}^{22}
\end{bmatrix}
c_w^{(i)}
= 
\begin{bmatrix}
	\left( {\cal Q}_{11}^{2} \right)^T V_{H_{11}} \\
	\left( {\cal Q}_{21}^{2} \right)^T V_{H_{21}} \\
	\vdots \\
	\left( {\cal Q}_{\bar{q}\bar{m}}^{2} \right)^T V_{H_{\bar{q}\bar{m}}} \\			
\end{bmatrix}
\,,
\label{eq:vf_fastvf2}
\end{equation}
which is solved in least squares sense to determine $c_w^{(i)}$, and compute the new poles estimate with~\eqref{eq:vf_matrix4poles}. Computational savings arise from the fact that the size of the matrices involved in~\eqref{eq:vf_fastvf1} and~\eqref{eq:vf_fastvf2} is much lower than the size of the coefficient matrix in~\eqref{eq:vf_sysmimo}. Furthermore, since the $\bar{q} \bar{m}$ QR decompositions~\eqref{eq:vf_fastvf1} are independent, they can be performed in parallel~\cite{chinea2011parallelization}. The Fast VF algorithm with parallelization can identify reduced models for systems with hundreds of inputs and outputs  in minutes~\cite{grivet2015passive}. A pseudocode of a real-valued implementation of the Fast VF algorithm will be given in Sec.~\ref{sec:vf_real}.

Several other ideas were proposed to increase VF scalability for large input and output counts. In VF with compression, samples $H_k$ are ``compressed'' with a singular value decomposition reducing the cost of the subsequent fitting~\cite{grivet2011compression} and passivity enforcement steps~\cite{olivadese2012compressed}. The Loewner method~\cite{lefteriu2010new,kassis2016passive}, which is an alternative to VF for the data-driven modeling of linear systems, was also shown to scale favorably with respect to the number of inputs and outputs. This class of techniques is the subject of chapter~\ref{}.

\subsection{Example: modeling of a multiport interconnect on a printed circuit board}
\label{sec:vf_ex3}

\begin{figure}

	\begin{center}
		\includegraphics[width=0.8\linewidth]{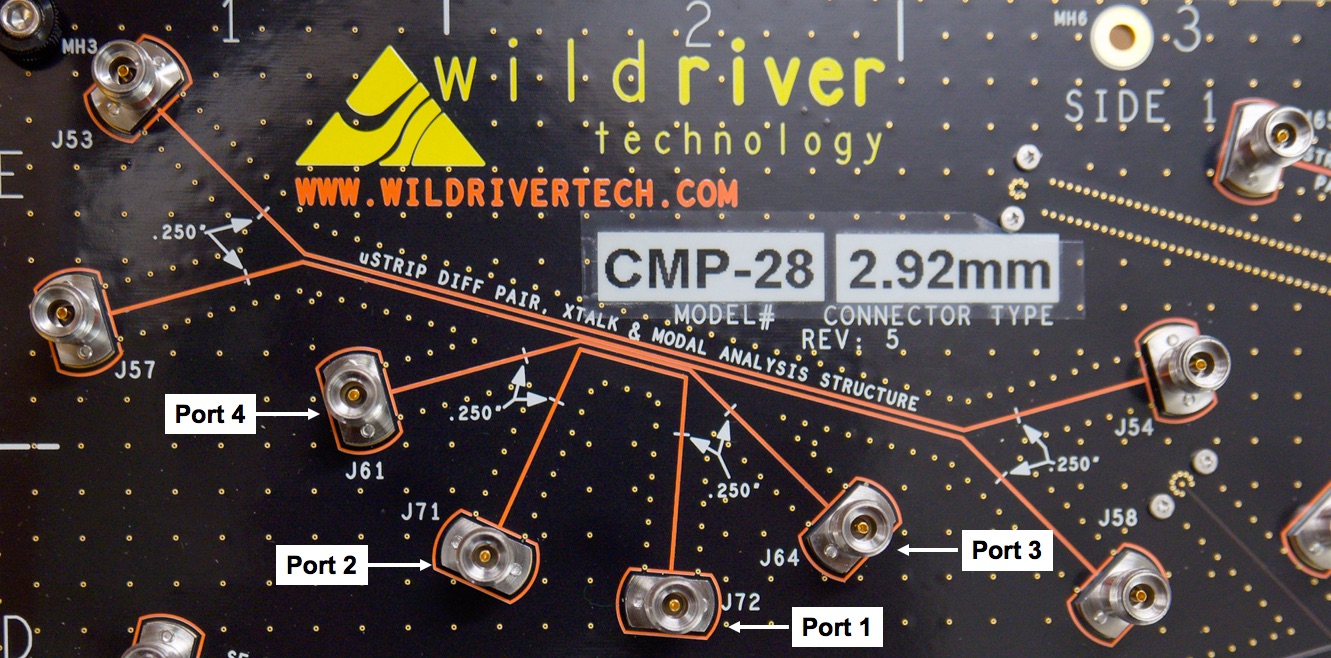}
	\end{center}
	\caption{Interconnect network on a printed circuit board considered in Sec.~\ref{sec:vf_ex3}. The four measurement ports of the vector network analyzer were connected as shown in the Figure.}
	\label{fig:vf_cmp28}
\end{figure}

Vector Fitting is extensively used by electronic designers to model how high-speed digital signals propagate over a printed circuit board, and design the system accordingly. We consider the structure shown in Fig.~\ref{fig:vf_cmp28}, which consists of several copper traces realized on the top face of a high-performance printed circuit board (Wild River Technology CMP-28~\cite{cmp28}). This structure mimics, in a simplified way, the multiwire buses that may connect the CPU and memory of a high-performance server. At the end of each trace, an electrical port is defined between the trace endpoint and a reference point on the ground plane underneath. The port is defined where the CPU or memory chip would be connected. In the test system, a high-frequency connector was installed at each port allowing the user to inject a signal from each port, and observe the signal received at the other ports.

In this example, we consider the two lower traces in Fig.~\ref{fig:vf_cmp28}, which have connectors J72, J71, J64, J61 soldered at their ends. The scattering matrix $H(\jmath \omega)$ of this 4-port device was measured from 10~MHz to 40~GHz in steps of 10~MHz with a Keysight N5227A vector network analyzer (courtesy of Fadime Bekmambetova, University of Toronto). In the scattering representation, input $U_m(\jmath \omega)$ is the amplitude of the electromagnetic wave injected into port $m$ by the instrument. Output $Y_q(\jmath \omega)$ is the amplitude of the wave received at port $q$. The scattering representation is commonly used at high frequency since it can be measured more accurately compared to the impedance or admittance representations used at low frequency.

\begin{figure}
	\centering
	\hspace*{3.5mm}\includegraphics[width=0.92\linewidth]{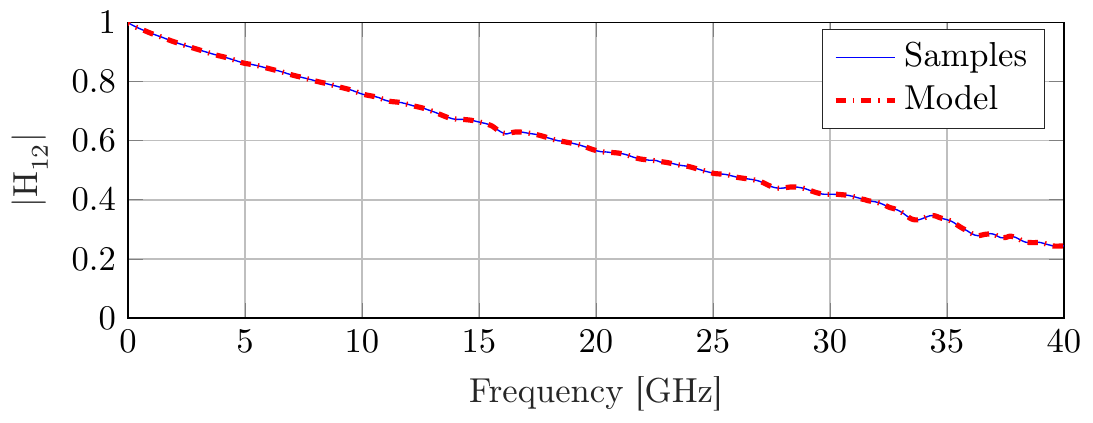}
	\includegraphics[width=0.94\linewidth]{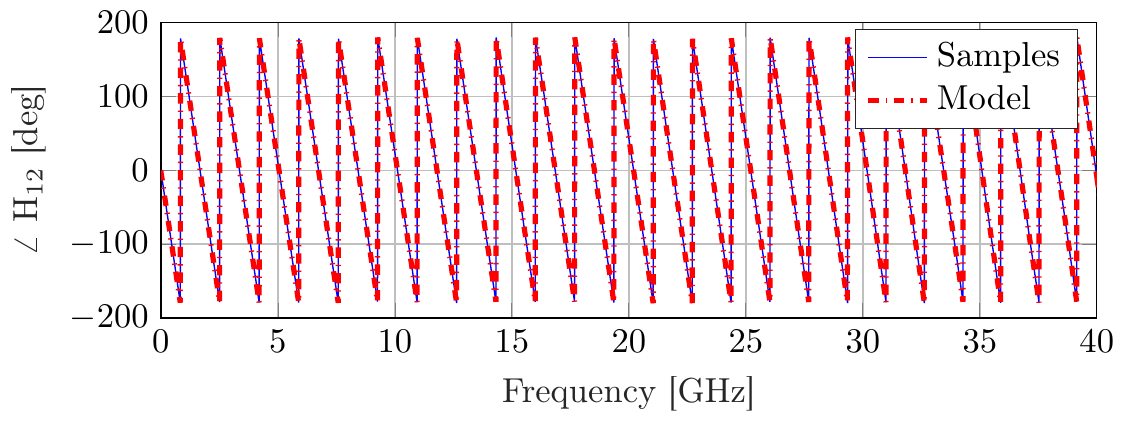}
	\caption{example of Sec.~\ref{sec:vf_ex3}: comparison between samples $H_{k,12}$ and corresponding VF model response.}
	\label{fig:vf_ex3S12}
		\begin{center}
		\hspace*{5.5mm}\includegraphics[width=0.9\linewidth]{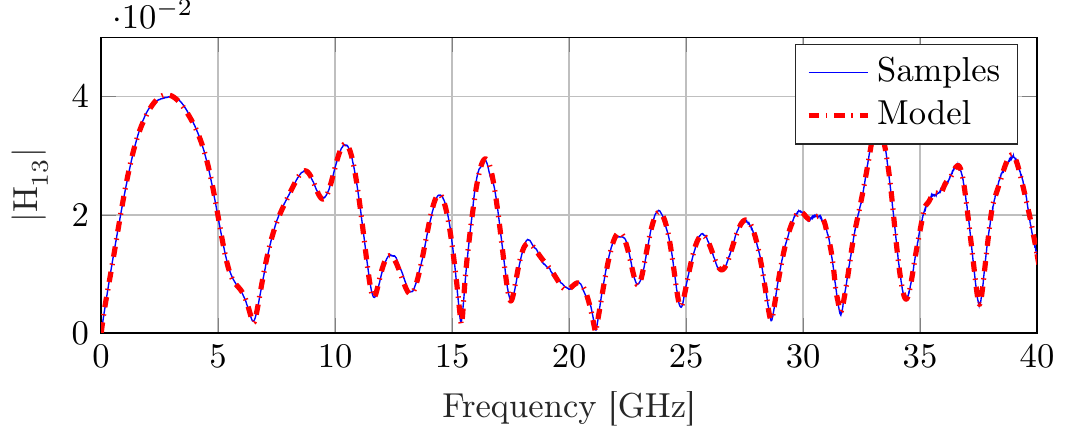}
		\includegraphics[width=0.94\linewidth]{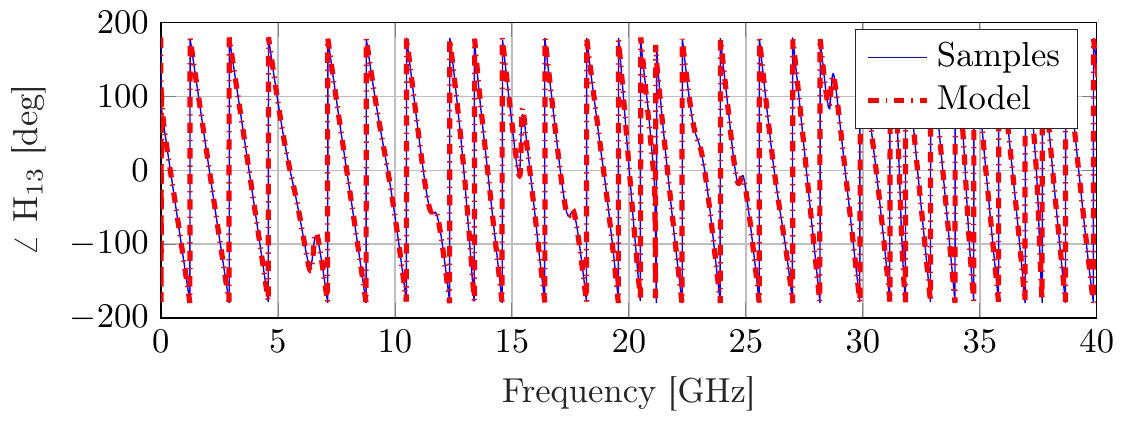}
	\end{center}
	\caption{example of Sec.~\ref{sec:vf_ex3}: comparison between samples $H_{k,13}$ and corresponding VF model response.}
	\label{fig:vf_ex3S13}
\end{figure}

A commercial implementation of the VF algorithm (IdEM, Dassault Systemes) was used to generate a reduced-order model from the measured samples (courtesy of Prof.~Stefano Grivet-Talocia, Politecnico di Torino). Figures~\ref{fig:vf_ex3S12} compares the VF model response to the original samples for the $(1,2)$ element of the scattering matrix. This response is the ratio between the amplitude of the wave received at one end of the trace (port~1) and the amplitude of the wave injected at the other end (port~2). We see that, as frequency increases, the received signal is progressively weaker, due to higher attenuation. The agreement between the VF model and the samples is excellent over the entire frequency range spanned by the measured data. Figure~\ref{fig:vf_ex3S13} compares the model response to the measured samples for the $(1,3)$ entry of the scattering matrix, which describes the signal received on the lower copper trace in Fig.~\ref{fig:vf_cmp28} when only the upper trace is excited. This coefficient is about 25~times smaller than the $(1,2)$ coefficient, since the two traces are not directly connected, and any coupling is due to electromagnetic interaction. We can see that the VF model approximates this small entry very accurately.  

\begin{figure}
	\begin{center}
		\includegraphics[width=0.9\linewidth]{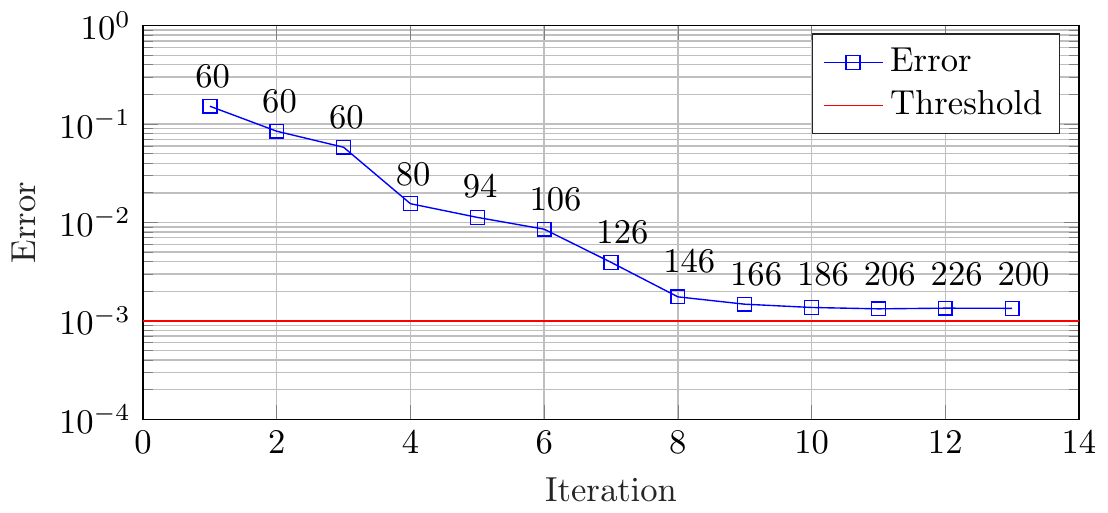}
	\end{center}
	\caption{example of Sec.~\ref{sec:vf_ex3}: VF error as a function of iteration, compared to the desired error level. Labels indicate the order $\bar{n}$ used by VF at each iteration.}
	\label{fig:vf_ex3_error}
\end{figure}

Figure~\ref{fig:vf_ex3_error} plots the samples-model error $e_{SK}^{(i)}$ as a function of $i$, together with the order $\bar{n}$ used by VF at each iteration. In this example, the order is adapted throughout iterations with the adding and skimming process~\cite{grivet2006improving} described in Sec.~\ref{sec:vf_orderestimation}. We observe that VF is able to progressively reduce the error throughout iterations, but convergence is slower than in the analytical example of Sec.~\ref{sec:vf_ex1}. This happens because of two reasons. First, this implementation of VF adaptively determines  order $\bar{n}$ in a single run, without requiring the user to determine a suitable $\bar{n}$ with multiple VF runs. Second, some noise is unavoidably present in the experimental measurements, which slows down convergence, and prevents VF from reducing the fitting error below $10^{-3}$. Indeed, we can see that VF is unable to increase model accuracy after the 10th iteration. Ultimately, VF delivers a reduced model with an error of $1.34\cdot 10^{-3}$, which is adequate for most design purposes.

\subsection{A real-valued formulation of VF and fast VF}
\label{sec:vf_real}

In most systems of practical interest, input $u(t)$ and output $y(t)$ are real-valued. Consequently, poles $p_n$ and residues $R_n$ are expected to be either real or in complex conjugate pairs. Because of round-off errors, the VF algorithm described so far may not ensure this realness condition. In this section, we describe a real-valued version of Fast VF which can be implemented in real arithmetics, and will ensure the realness condition by construction. The pseudo-code of the described algorithm is given in Algorithm~\ref{alg:vf_fastvfreal}. An open-source implementation of this algorithm, which closely follows the notation and pseudocode in this chapter, can be downloaded from~\cite{myvfwebsite}.

To ensure complex conjugate poles and residues, we redefine model~\eqref{eq:vf_Hmimo} as
\begin{equation}
	\widetilde{H}^{(i)}(s) 	= R_0^{(i)} +  \sum_{n=1}^{\bar{n}_r} \frac{R_n^{(i)}}{s-p_n^{(i-1)}}
		+  \sum_{n=\bar{n}_r+1}^{\bar{n}_r+\bar{n}_c} \left[ \frac{R_{n}^{(i)}}{s-p_{n}^{(i-1)}} + 
		\frac{\left(R_{n}^{(i)}\right)^*}{s-\left(p_{n}^{(i-1)}\right)^*} \right]	
	\,,
	\label{eq:vf_Hmimoreal}
\end{equation}
where $\bar{n}_r$ is the number of real poles and $\bar{n}_c$ is the number of pairs of complex conjugate poles, for a total order $\bar{n} = \bar{n}_r + 2 \bar{n}_c$. In~\eqref{eq:vf_Hmimoreal}, we force $R_n^{(i)} \in \mathbb{R}$ for $n=0,\dots, \bar{n}_r$. The VF weighting function~\eqref{eq:vf_w} is redefined in a similar fashion as
\begin{equation}
	w^{(i)}(s) = 1 +  \sum_{n=1}^{\bar{n}_r} \frac{w_n^{(i)}}{s-p_n^{(i-1)}}
	+  \sum_{n=\bar{n}_r+1}^{\bar{n}_r+\bar{n}_c} \left[ \frac{w_{n}^{(i)}}{s-p_{n}^{(i-1)}} + 
	\frac{\left(w_{n}^{(i)}\right)^*}{s-\left(p_{n}^{(i-1)}\right)^*} \right]	
	\,,
	\label{eq:vf_wreal}
\end{equation}
where $w_n^{(i)} \in \mathbb{R}$ for $n=1,\dots,\bar{n}_r$. Using~\eqref{eq:vf_Hmimoreal} and~\eqref{eq:vf_wreal}, and following the steps in Sec.~\ref{sec:vf_mimo}, one can arrive at a least-squares system in the same form as~\eqref{eq:vf_sysmimo}
\begin{equation}
	\begin{bmatrix}
		\Phi_0^{(i)}	&	0 & \dots & 0  &-D_{H_{11}} \Phi_1^{(i)} \\
		0					&		\Phi_0^{(i)} & \ddots & \vdots & -D_{H_{21}} \Phi_1^{(i)} \\
		\vdots & \ddots & \ddots & 0 & \vdots\\
		0 & \dots & 0 & \Phi_0^{(i)} & -D_{H_{\bar{q} \bar{m}}} \Phi_1^{(i)} \\
	\end{bmatrix}
	\begin{bmatrix}
		c_{H_{11}}^{(i)} \\		
		c_{H_{21}}^{(i)} \\
		\vdots \\
		c_{H_{\bar{q} \bar{m}}}^{(i)} \\
		c_w^{(i)}\\
	\end{bmatrix}
	=
	\begin{bmatrix}
		V_{H_{11}} \\
		V_{H_{21}} \\
		\vdots \\
		V_{H_{\bar{q} \bar{m}}}
	\end{bmatrix}\,,
	\label{eq:vf_sysmimo2}
\end{equation}
but where we take the real and imaginary part of each residue in~\eqref{eq:vf_Hmimoreal} as unknowns
\begin{equation}
	c_{H_{qm}}^{(i)} = 
	\begin{bmatrix}
		R_{0,qm}^{(i)} & \dots & R_{\bar{n}_r,qm}^{(i)} &
		\vfRe{R_{\bar{n}_r+1,qm}^{(i)}} & \vfIm{R_{\bar{n}_r+1,qm}^{(i)}} & \dots
	\end{bmatrix}^T\,,
	\label{eq:vf_cHqmreal}
\end{equation}
and the real and imaginary part of each residue of the weighting function~\eqref{eq:vf_wreal}
\begin{equation}
	c_{w}^{(i)} = 
	\begin{bmatrix}
		w_{1}^{(i)} & \dots & w_{\bar{n}_r}^{(i)}  &
		\vfRe{w_{\bar{n}_r+1}^{(i)}} & \vfIm{w_{\bar{n}_r+1}^{(i)}} & \dots
	\end{bmatrix}^T\,.
	\label{eq:vf_cwreal}
\end{equation}
This choice of unknowns will ensure that complex residues always come in conjugate pairs. The coefficient matrices $\Phi_0^{(i)}$ and $\Phi_1^{(i)}$ in~\eqref{eq:vf_sysmimo2} are given by
\begin{align}
	\Phi_0^{(i)} & = 
	\begin{bmatrix}
		1_{\bar{k}} & \Phi_r^{(i)} & \Phi_{c}^{(i)} 
	\end{bmatrix}\,, \\
	\Phi_1^{(i)} & = 
	\begin{bmatrix}
		\Phi_r^{(i)} &  \Phi_{c}^{(i)} 
	\end{bmatrix}\,,
\end{align}
where $1_{\bar{k}}$ is a $\bar{k} \times 1$ vector of ones, and
\begin{align}
	\Phi_r^{(i)} & = 
	\begin{bmatrix}
		\frac{1}{\jmath \omega_1 - p_1^{(i-1)}} & \dots & \frac{1}{\jmath \omega_1 - p_{\bar{n}_r}^{(i-1)}}  \\
		\vdots & \ddots & \vdots \\
		\frac{1}{\jmath \omega_{\bar{k}} - p_1^{(i-1)}} & \dots & \frac{1}{\jmath \omega_{\bar{k}} - p_{\bar{n}_r}^{(i-1)}} \\
	\end{bmatrix}\,,
	\label{eq:vf_Phir}\\
	\Phi_{c}^{(i)} & = 
\begin{bmatrix}
	\frac{1}{\jmath \omega_1 - p_{\bar{n}_r+1}^{(i-1)}} + \frac{1}{\jmath \omega_1 - \left(p_{\bar{n}_r+1}^{(i-1)}\right)^*}
	 & \frac{\jmath}{\jmath \omega_1 - p_{\bar{n}_r+1}^{(i-1)}} - \frac{\jmath}{\jmath \omega_1 - \left(p_{\bar{n}_r+1}^{(i-1)}\right)^*} 
	 & \dots \\
	\vdots & \vdots & \\
	\frac{1}{\jmath \omega_{\bar{k}} - p_{\bar{n}_r+1}^{(i-1)}} + \frac{1}{\jmath \omega_{\bar{k}} - \left(p_{\bar{n}_r+1}^{(i-1)}\right)^*}
& \frac{\jmath}{\jmath \omega_{\bar{k}} - p_{\bar{n}_r+1}^{(i-1)}} - \frac{\jmath}{\jmath \omega_{\bar{k}} - \left(p_{\bar{n}_r+1}^{(i-1)}\right)^*} & \dots
\end{bmatrix}\,,
	\label{eq:vf_Phic} 
\end{align}
Although~\eqref{eq:vf_sysmimo2} has real unknowns, its coefficients matrix and right hand side are still complex-valued. To remedy this issue, we write the real and imaginary part of each equation separately
\begin{equation}
	\begin{bmatrix}
	\vfRe{\Phi_0^{(i)}}	&	\!\!0 & \!\!\dots & 0  & \!\!\!\!-\vfRe{D_{H_{11}} \Phi_1^{(i)}} \\
	\vfIm{\Phi_0^{(i)}}	&	\!\!0 & \!\!\dots & 0  & \!\!\!\!-\vfIm{D_{H_{11}} \Phi_1^{(i)}} \\	
	\vdots &  &  & \vdots & \vdots\\
	0 & \!\!\dots & \!\!0 & \vfRe{\Phi_0^{(i)}} & \!\!\!\!-\vfRe{D_{H_{\bar{q} \bar{m}}} \Phi_1^{(i)}} \\
	0 &\!\! \dots & \!\!0 & \vfIm{\Phi_0^{(i)}} & \!\!\!\!-\vfIm{D_{H_{\bar{q} \bar{m}}} \Phi_1^{(i)}} \\	
	\end{bmatrix} \!\!\!\!
	\begin{bmatrix}
	c_{H_{11}}^{(i)} \\		
	\vdots \\
	c_{H_{\bar{q} \bar{m}}}^{(i)} \\
	c_w^{(i)}\\
	\end{bmatrix}
	\!\!=\!\!
	\begin{bmatrix}
	\vfRe{V_{H_{11}}} \\
	\vfIm{V_{H_{11}}} \\
	\vdots \\
	\vfRe{V_{H_{\bar{q} \bar{m}}}} \\
	\vfIm{V_{H_{\bar{q} \bar{m}}}}	
	\end{bmatrix}\,.
	\label{eq:vf_sysmimoreal}
\end{equation}
The obtained system, which has real coefficients and unknowns will ensure, by construction, that model poles and residues are either real or complex conjugate. Due to its block structure, system~\eqref{eq:vf_sysmimoreal} can be efficiently solved with the Fast VF approach discussed in Sec.~\ref{sec:vf_fastvf}. In step~\ref{alg:vf_fast_fit1} of Algorithm~\ref{alg:vf_fastvfreal}, the QR decompositions
\begin{equation}
	\begin{bmatrix}
		\vfRe{\Phi_0^{(i)}}  & -\vfRe{D_{H_{qm}} \Phi_1^{(i)}} \\
		\vfIm{\Phi_0^{(i)}}  & -\vfIm{D_{H_{qm}} \Phi_1^{(i)}}		
	\end{bmatrix}
	= 
	\begin{bmatrix}
		{\cal Q}_{qm}^{11} & {\cal Q}_{qm}^{12} \\
		{\cal Q}_{qm}^{21} & {\cal Q}_{qm}^{22}		
	\end{bmatrix}
	\begin{bmatrix}
		{\cal R}_{qm}^{11}	& 	{\cal R}_{qm}^{12} \\
		0						& 	{\cal R}_{qm}^{22}
	\end{bmatrix}
	\,,
\label{eq:vf_fastvf1real}
\end{equation}
are computed for $q=1,\dots,\bar{q}$ and $m=1,\dots,\bar{m}$. Then, in step~\ref{alg:vf_fast_fit1}, reduced system
\begin{equation}
	\begin{bmatrix}
		{\cal R}_{11}^{22} \\
		{\cal R}_{21}^{22} \\
		\vdots \\
		{\cal R}_{\bar{q}\bar{m}}^{22}
	\end{bmatrix}
	c_w^{(i)}
	= 
	\begin{bmatrix}
		\left( {\cal Q}_{11}^{12} \right)^T \vfRe{V_{H_{11}}} +\left( {\cal Q}_{11}^{22} \right)^T \vfIm{V_{H_{11}}} \\
		\left( {\cal Q}_{21}^{12} \right)^T \vfRe{V_{H_{21}}} +\left( {\cal Q}_{21}^{22} \right)^T \vfIm{V_{H_{21}}} \\
		\vdots \\
		\left( {\cal Q}_{\bar{q}\bar{m}}^{12} \right)^T \vfRe{V_{H_{\bar{q}\bar{m}}}} +\left( {\cal Q}_{\bar{q}\bar{m}}^{22} \right)^T \vfIm{V_{H_{\bar{q}\bar{m}}}} \\
	\end{bmatrix}
	\,,
	\label{eq:vf_fastvf2real}
\end{equation}
is solved in least squares sense to determine $c_w^{(i)}$ and compute the new poles estimate with the real-valued counterpart of~\eqref{eq:vf_matrix4poles}, which reads~\cite{grivet2015passive}
\begin{equation}
	\left\{ p_n^{(i)} \right\} = \text{eig} \left( A^{(i-1)}  - b_w  \left( c_w^{(i)} \right)^T \right) \,,
	\label{eq:vf_matrix4polesreal}
\end{equation}
with $A^{(i-1)} = \text{diag} \left \{ p_1^{(i-1)},\dots,p_{\bar{n}_r}^{(i-1)}, \Pi_{\bar{n}_r+1}^{(i-1)}, \dots, \Pi_{\bar{n}_r+\bar{n}_c}^{(i-1)} \right \}$ being a block diagonal matrix formed by the real poles and, for complex conjugate pairs, by the blocks
\begin{equation}
	\Pi_n^{(i-1)} =
	\begin{bmatrix}
		\vfRe{p_n^{(i-1)}} & \vfIm{p_n^{(i-1)}} \\
		-\vfIm{p_n^{(i-1)}} & \vfRe{p_n^{(i-1)}} 
	\end{bmatrix}\,.
\end{equation}
In~\eqref{eq:vf_matrix4polesreal}, $b_w$ is a $\bar{n} \times 1$ vector with the first $\bar{n}_r$ entries set to one, followed by a $[2,0]^T$ block for each pair of complex conjugate poles.

Once poles have been estimated, a first convergence test is performed in step~\ref{alg:vf_fast_conv1} using~\ref{eq:vf_convw}. If the test is passed, in step~\ref{alg:vf_fast_fit2} of Algorithm~\ref{alg:vf_fastvfreal} we fit the residues of the final model, solving in least squares sense
\begin{equation}
	\begin{bmatrix}
		\vfRe{\Phi_0^{(i+1)}}  \\
		\vfIm{\Phi_0^{(i+1)}} 
	\end{bmatrix}
	c_{H_{qm}}^{(i+1)} =
	\begin{bmatrix}
		\vfRe{V_{H_{qm}}} \\
		\vfIm{V_{H_{qm}}}
	\end{bmatrix}\,,
	\label{eq:vf_sysfixedpolesmimoreal}
\end{equation} 
for $q=1,\dots,\bar{q}$ and $m=1,\dots,\bar{m}$. The second and final convergence test is performed in step~\ref{alg:vf_fast_conv2} of Algorithm~\ref{alg:vf_fastvfreal}.

\begin{algorithm}[t]
	\caption{Fast Vector Fitting, real-valued implementation}
	\label{alg:vf_fastvfreal}
	\begin{algorithmic}[1]
		\Require response samples $H_k$, corresponding frequencies $\omega_k$ ($k=1,\dots,\bar{k})$
		\Require desired model order $\bar{n}$
		\Require maximum number of iterations $i_{max}$
		\State set initial poles $p_n^{(0)}$ according to~\eqref{eq:vf_p0DC} or~\eqref{eq:vf_p0}.
		\State $i \gets 1$
		\While{$i \le i_{max}$}
		\State Compute QR decompositions~\eqref{eq:vf_fastvf1real} \label{alg:vf_fast_qr}
		\State Solve~\eqref{eq:vf_fastvf2real} in least squares sense \label{alg:vf_fast_fit1}
		\State Compute the new poles estimate $p_n^{(i)}$ with~\eqref{eq:vf_matrix4polesreal} \label{alg:vf_fast_np}
		\State Enforce poles stability with~\eqref{eq:vf_stablepoles}, if desired \Comment Stability enforcement
		\If{\eqref{eq:vf_convw} is true} \Comment First convergence test \label{alg:vf_fast_conv1}
		\State Solve~\eqref{eq:vf_sysfixedpolesmimoreal} in least squares sense \label{alg:vf_fast_fit2} \Comment Tentative final fitting
		\State Compute fitting error $e$ with~\eqref{eq:vf_emimo}
		\If{$e \le \varepsilon_H$} \Comment Second convergence test \label{alg:vf_fast_conv2}
		\State $\widetilde{H}(s) = \widetilde{H}^{(i+1)}(s)$
		\State \Return Success!
		\EndIf
		\EndIf
		\State $i \gets i+1$
		\EndWhile
		\State \Return Failure: maximum number of iterations reached.
	\end{algorithmic}
\end{algorithm}
\clearpage

\subsection{Model realization}
\label{sec:vf_realization}

The real-valued formulation of VF, discussed in Sec.~\ref{sec:vf_real}, produces a reduced model in the form
\begin{equation}
	\widetilde{H}(s) 	= R_0 +  \sum_{n=1}^{\bar{n}_r} \frac{R_n}{s-p_n} + \sum_{n=\bar{n}_r+1}^{\bar{n}_r+\bar{n}_c} 
	\left[
		\frac{R_n}{s-p_n}  + \frac{R^*_n}{s-p^*_n}
	\right] \,,
\label{eq:vf_Hmimofinal}
\end{equation}
which can be easily converted into a variety of equivalent representations to facilitate its use in different simulation scenarios. Expression~\eqref{eq:vf_Hmimofinal} is known as pole-residue form of the transfer function. This form is the most convenient when the model will be used in frequency-domain analyses, since it minimizes the computational cost of evaluating $H(\jmath \omega)$. 

For time-domain analyses, such as transient simulations, expression~\eqref{eq:vf_Hmimofinal} can be converted into the time-domain with the inverse Laplace transform, which yields
\begin{equation}
	\widetilde{h}(t) 	= R_0 +  \sum_{n=1}^{\bar{n}_r} R_n e^{p_n t}  +
	\sum_{n=\bar{n}_r+1}^{\bar{n}_r+\bar{n}_c} 
	\left[
		2 R'_n e^{p'_n t} \cos (p''_n t) - 2 R''_n e^{p'_n t} \sin (p''_nt) 
	\right]
	\label{eq:vf_hmimo}
\end{equation}
for $t \ge 0$, where $p'_n  = \vfRe{p_n}$, $p''_n  = \vfIm{p_n}$, $R'_n  = \vfRe{R_n}$ and $R''_n  = \vfIm{R_n}$. In~\eqref{eq:vf_hmimo}, $\widetilde{h}(t)$ denotes the impulse response of the model. This form is particularly convenient in transient simulators based on convolutions like~\eqref{eq:vf_systd}. While computing convolution integrals is in general very expensive, when an impulse response has the form~\eqref{eq:vf_hmimo}, convolution can be computed very quickly using recursive formulas~\cite{grivet2015passive}. 

While convolutional simulators are prominent in selected applications, the majority of transient simulators is based on the solution of differential equations, and cannot handle~\eqref{eq:vf_hmimo} directly. To overcome this issue, we can represent~\eqref{eq:vf_Hmimofinal} through a set of differential equations in state-space form
\begin{equation}
	\begin{cases}
		\dot{x}(t) = A x (t) + B u(t)\,, \\
		y(t) = C x(t) + D u(t)\,.
	\end{cases}
	\label{eq:vf_ss}
\end{equation}
System~\eqref{eq:vf_ss} is constructed in such a way that the transfer function between input $u(t)$ and output $y(t)$ is~\eqref{eq:vf_Hmimofinal}. Given a transfer function, there are infinitely-many systems~\eqref{eq:vf_ss} that meet this criterion, known as \emph{realizations} of $H(s)$. We present a popular realization, due to Gilbert~\cite{gilbert1963controllability}, and refer the Reader to~\cite{grivet2015passive} for a comprehensive description of how VF models can be realized.

For reasons that will become clear later on, the Gilbert realization process begins with the truncated singular value decomposition~\cite{Gol96} of residues $R_n$
\begin{equation}
	R_n = U_n \Sigma_n V_n^H \qquad  \text{for } n=1,\dots,\bar{n}_r+\bar{n}_c \,,
	\label{eq:vf_svd}
\end{equation}
where $\Sigma_n = \text{diag}\{ \sigma_{n,1}, \dots, \sigma_{n,\rho_n}\}$ is a diagonal matrix collecting all nonzero singular values of $R_n$, and $\rho_n$ is the rank of $R_n$. Matrices $U_n \in \mathbb{C}^{\bar{q} \times \rho_n}$ and $V_n \in \mathbb{C}^{\bar{m} \times \rho_n}$ are formed by the left and right singular vectors of $R_n$, respectively. In~\eqref{eq:vf_svd}, $^H$ denotes the conjugate transpose, also known as Hermitian transpose. Given~\eqref{eq:vf_svd}, we can express the partial fractions in~\eqref{eq:vf_Hmimofinal} associated to real poles as
\begin{equation}
	\frac{R_n}{s-p_n} = U_n \Sigma_n \frac{I_{\rho_n}}{s-p_n} V_n^T = C_n \left(s I_{\rho_n} - A_n\right)^{-1} B_n\,,
	\label{eq:vf_Rntf_real}
\end{equation}
for $n=1,\dots, \bar{n}_r$. In~\eqref{eq:vf_Rntf_real},  $I_{\rho_n}$ is the identity matrix of size $\rho_n \times \rho_n$, $C_n = U_n \Sigma_n$, $A_n = p_n I_{\rho_n}$, and $B_n = V_n^T$. For complex poles, we can derive an equivalent expression for the sum of the two conjugate partial fractions~\cite{grivet2015passive}
\begin{equation}
	\frac{R_n}{s-p_n} + \frac{R^*_n}{s-p^*_n}= C_n \left(s I_{2\rho_n} - A_n\right)^{-1} B_n\,,
	\label{eq:vf_Rntf_complex}
\end{equation}
for $n=\bar{n}_r+1,\dots, \bar{n}_r + \bar{n}_c$, where
\begin{align}
	A_n = & 
	\begin{bmatrix} 
		p'_n I_{\rho_n} 	& p''_n I_{\rho_n} \\
		-p''_n I_{\rho_n} 	& p'_n I_{\rho_n}
	\end{bmatrix}  &
	B_n = &  2
	\begin{bmatrix}
		 \vfRe{V_n^T} \\
		 \vfIm{V_n^T}
	\end{bmatrix} \\
	C_n =&  
	\begin{bmatrix}
		\vfRe{U_n \Sigma_n} & \vfIm{U_n\Sigma_n}
	\end{bmatrix}\,. 
\end{align}
 Expressions~\eqref{eq:vf_Rntf_real} and~\eqref{eq:vf_Rntf_complex} allow us to rewrite~\eqref{eq:vf_Hmimofinal} as
\begin{equation}
	\widetilde{H}(s) = D + C \left(s I_{N} - A \right)^{-1} B\,,
	\label{eq:vf_Hss}
\end{equation}
where
\begin{align}
	A = & \begin{bmatrix} A_1 & & \\
											& \ddots & \\
											& & A_{\bar{n}_r+\bar{n}_c}
			\end{bmatrix}  &
	B = & \begin{bmatrix}
		B_1 \\
		\vdots \\
		B_{\bar{n}_r+\bar{n}_c}
	\end{bmatrix} \label{eq:vf_AB}\\
	C =&  \begin{bmatrix}
		C_1 & \dots & C_{\bar{n}_r+\bar{n}_c}
	\end{bmatrix}  &
	D = & R_0\,.  \label{eq:vf_CD}
\end{align}
Since~\eqref{eq:vf_Hss} is the transfer function of~\eqref{eq:vf_ss}, equations~\eqref{eq:vf_AB} and~\eqref{eq:vf_CD} provide the coefficient matrices of a state space realization~\eqref{eq:vf_ss} of transfer function~\eqref{eq:vf_Hmimofinal} produced by VF. The order of~\eqref{eq:vf_Hss} is 
\begin{equation}
	N = \sum_{n=1}^{\bar{n}_r} \rho_n + 2\sum_{n=\bar{n}_r+1}^{\bar{n}_r+\bar{n}_c} \rho_n
	\label{eq:vf_order}
\end{equation}
and can be shown to be minimal~\cite{gilbert1963controllability}. This property stems from the singular value decompositions~\eqref{eq:vf_svd}, which reveal the rank $\rho_n$ of each residue $R_n$. If those singular value decompositions are not performed, a realization of order $\bar{n} \bar{m}$ is obtained. This realization may not be minimal, and may contain states that are not controllable, not observable, or both, as discussed in chapter~\ref{}.

In addition to the forms presented in this section, the VF model~\eqref{eq:vf_Hmimofinal} can be converted to a variety of additional forms, including equivalent electric circuits~\cite{antonini2003spice,grivet2015passive} for seamless integration into any circuit simulator.

\subsection{Stability, causality and passivity enforcement}

Most systems of practical interest are stable, and the real part of their poles is either negative or zero. One would expect that, given noise-free samples of their frequency response, VF will produce a model with stable poles satisfying
\begin{equation}
	\text{Re} \{ p_n\} \le 0 \quad \forall n\,.
	\label{eq:vf_stable}
\end{equation}
Unfortunately, this is not guaranteed, since round-off errors may indeed push a few poles into the right half of the complex plane, making the VF model unstable.

Condition~\eqref{eq:vf_stable} is essentially mandatory for time-domain simulations, since otherwise results will diverge. The standard practice is to enforce stability during VF iterations. After computing the new poles estimate $p_n^{(i)}$ with~\eqref{eq:vf_matrix4poles}, the following rule is applied
\begin{equation}
	p_n^{(i)} = 
	\begin{cases}
		p_n^{(i)} & \text{ if } \vfRe{p_n^{(i)}} < 0\,,\\
		-\vfRe{p_n^{(i)} } + \jmath \vfIm{p_n^{(i)} } & \text{ if } \vfRe{p_n^{(i)}} > 0\,,
	\end{cases}
	\label{eq:vf_stablepoles}
\end{equation}
for $n= 1, \dots, \bar{n}_r + \bar{n}_c$. We can see that, if a pole $p_n^{(i)}$ is unstable, the sign of its real part is inverted. Since in the tentative final fitting in step~\ref{alg:vf_fit2} of Algorithm~\ref{alg:vf_vf} poles are fixed, condition~\eqref{eq:vf_stablepoles} ensures the stability of the final model.

For frequency domain analyses, one may think that~\eqref{eq:vf_stable} is not necessary, since stability is not an issue. However, one can show that~\eqref{eq:vf_stable}, in the frequency domain, becomes a condition for causality~\cite{triverio2007stability}. Causality means that the system will react to an excitation only after it has been applied, and not before. In other words, if the system input $u(t)$ begins at $t=t_0$ ($u(t)=0$ for $t < t_0$), the system output will start varying only at or after $t=t_0$. All systems in nature are obviously causal, since they cannot ``anticipate'' the application of an excitation. Enforcing~\eqref{eq:vf_stable} ensures that VF model~\eqref{eq:vf_Hmimofinal} is causal. If this is not the case, frequency-domain analyses will succeed, but results may be inaccurate and unphysical. In particular, the VF model may underestimate the delay between input and output which is present in the real system, which may be important in some applications, such as the timing analysis of digital circuits. A complete discussion of causality is beyond the scope of this chapter, and the Reader is referred to~\cite{triverio2007stability}. 

Overall, condition~\eqref{eq:vf_stable} simultaneously enforces the stability and causality of the VF model. This condition can be enforced without any accuracy penalty when the given samples $H_k$ are error free, and thus faithfully represent the response of a causal and stable system. When samples are corrupted by noise or measurement errors, VF may be unable to reduce fitting error~\eqref{eq:vf_e} to the desired level if condition~\eqref{eq:vf_stable} is enforced. This happens when the noise or errors in samples $H_k$ are not causal functions themselves, and thus cannot be approximated with stable and causal poles~\cite{triverio2007stability}. Numerical algorithms exist to verify if the given samples $H_k$ satisfy the causality condition required by VF to fit them with high accuracy~\cite{triverio2006robust,triverio2008robust,lalgudi2013checking,triverio2014robust,barannyk2015spectrally}.

In addition to causality and stability, passivity is another important property that one may want to impose on the VF model~\eqref{eq:vf_Hmimofinal}. This property characterizes those physical systems that are unable to generate energy on their own, simply due to the lack of energy sources or gain mechanisms inside them. A circuit made by positive resistors, capacitors and inductors is an example of a passive system, in contrast to an amplifying circuit. When applied to the response of a passive system, VF may still produce a non-passive model, due to approximation and numerical errors. However, passivity can be enforced a-posteriori, with the methods presented in chapter~\ref{}.

\subsection{Numerical implementation}

Vector Fitting is easy to implement, and several free codes are available~\cite{myvfwebsite,gustavsen2013vectorfittingwebsite}. This section briefly describes a few changes to the basic templates in Algorithms~\ref{alg:vf_vf} and~\ref{alg:vf_fastvfreal} that can lead to a more robust and efficient implementation.

\subsubsection{Order estimation}
\label{sec:vf_orderestimation}

The VF templates in Algorithms~\ref{alg:vf_vf} and~\ref{alg:vf_fastvfreal} require the desired model order $\bar{n}$ as input. Typically, this is not known a priori, but can be determined during the fitting process using the VF algorithm with adding and skimming~\cite{grivet2006improving}, as shown in the example of Sec.~\ref{sec:vf_ex3}. In this method, an initial estimate of $\bar{n}$ is derived from the phase of the given samples $H_k$, and used in the first VF iteration. Then, $\bar{n}$ is automatically increased or decreased based on the achieved error, as visible in Fig.~\ref{fig:vf_ex3_error}. If error $e$ is still too high, the order is increased until either VF converges or it becomes evident that no further error reduction can be achieved, as in the last four iterations in Fig.~\ref{fig:vf_ex3_error}. Conversely, when the algorithm detects that some partial fractions in~\eqref{eq:vf_Hmimofinal} give a negligible contribution over the frequency range of interest, order $\bar{n}$ is reduced at the next iteration by removing such terms. This happens in the 13th iteration of the example in Sec.~\ref{sec:vf_ex3}, where order is reduced from 226 to 200.

\subsubsection{Relaxed VF: a better normalization of the weighting function}
\label{sec:vf_relaxed}

In the original VF algorithm, the coefficients of weighting function~\eqref{eq:vf_w} are normalized such that $w^{(i)}(\jmath \omega) \to 1$ when $\omega \to \infty$. It can be shown that this normalization is not optimal, and can slow down VF convergence when samples $H_k$ are contaminated by noise. The relaxed VF algorithm~\cite{gustavsen2006improving} mitigates this issue by redefining the weighting function as
\begin{equation}
	w^{(i)}(s) 
	= w_0^{(i)} +  \sum_{n=1}^{\bar{n}} \frac{w_n^{(i)}}{s-p_n^{(i-1)}}
	\,,
	\label{eq:vf_wrel}
\end{equation}
where $w_0^{(i)}$ is now free to depart from one. With this change, the fitting equation~\eqref{eq:vf_eqmimo} becomes
\begin{equation}
	R_{0,qm}^{(i)} + \sum_{n=1}^{\bar{n}} \frac{R_{n,qm}^{(i)}}{\jmath \omega_k - p_n^{(i-1)}}
	- H_{k,qm} \left( w_0^{(i)} + \sum_{n=1}^{\bar{n}} \frac{w_n^{(i)}}{\jmath \omega_k-p_n^{(i-1)}} \right)
	= 0\,.
	\label{eq:vf_eqmimorel}
\end{equation}
Since~\eqref{eq:vf_eqmimorel} admits a trivial solution ($R_{n,qm}^{(i)} = w_n^{(i)} = 0$ $\forall n$), the relaxed VF algorithm adds an additional constraint to exclude it~\cite{gustavsen2006improving}
\begin{equation}
	\frac{1}{\bar{k}} \sum_{k=1}^{\bar{k}} \text{Re} \left \{ 
		w_0^{(i)} +  \sum_{n=1}^{\bar{n}} \frac{w_n^{(i)}}{\jmath \omega_k-p_n^{(i-1)}}
		\right \} = 1\,.
	\label{eq:vf_rel}
\end{equation}
This constraint can be seen as a more relaxed normalization of the weighting function. Equations~\eqref{eq:vf_eqmimorel} and~\eqref{eq:vf_rel} are then jointly solved in least squares sense. In the single-input single-output case ($\bar{q} = \bar{m} = 1$), the system to be solved takes the form

\begin{equation}
	\begin{bmatrix}
	\Phi_0^{(i)} 	& -D_H \Phi_0^{(i)} \\
	0 					  & \frac{\beta}{\bar{k}} \left(1_{\bar{k}}\right)^T \Phi_0^{(i)}
	\end{bmatrix}
	\begin{bmatrix}
	c^{(i)}_H \\ c^{(i)}_w
	\end{bmatrix}	
	= 
	\begin{bmatrix}
		0 \\
		\beta
	\end{bmatrix}
	\label{eq:vf_sysrel}
\end{equation}
where
\begin{equation}
	c_w^{(i)} = 
	\begin{bmatrix}
		w_0^{(i)} & \dots & w_{\bar{n}}^{(i)} 
	\end{bmatrix}^T\,.
	\label{eq:vf_cwrel}
\end{equation}
In~\eqref{eq:vf_sysrel}, $\beta$ is a suitable weight to the last equation, which is typically set to~\cite{gustavsen2006improving}
\begin{equation}
	\beta = \sqrt{\sum_{k=1}^{\bar{k}} \left| H_k \right|^2}\,.
	\label{eq:vf_beta}
\end{equation}

\section{Generalized and advanced VF algorithms}

Since its inception in 1996, VF has inspired a generation of algorithms for the data-driven modeling of linear systems. These extensions either improve the original VF formulation, or extend it to different modeling scenarios. We briefly summarize the most relevant works in this area, and provide several bibliographic references where more details can be found.

\subsection{Time-domain VF algorithms}

The original VF algorithm works in the frequency domain, and creates the reduced model from samples of the system frequency response. In some applications, however, it is more convenient to characterize the system in the time domain. For example, one may have simultaneous measurements of the system input $u(t_l)$ and output $y(t_l)$ at several time points $t_l$ for $l=1,\dots,\bar{l}$, as in the example of Sec.~\ref{sec:vf_ex2}. In this scenario, one has two options. The first is to estimate the systems' frequency response from the time-domain samples with the discrete Fourier transform, and apply VF in the frequency domain. However, the accuracy of the discrete Fourier transform depends significantly on the sampling rate of the given samples, and on their behavior near the boundaries $t=t_1$ and $t=t_{\bar{l}}$ of the acquisition window. These issues, if not well understood and managed, can result in an inaccurate time-frequency conversion, and degrade model quality.
	
The second option is to use the time-domain VF algorithm~\cite{grivet2003package,grivet2004time}, which directly extracts~\eqref{eq:vf_Hmimofinal} from the time-domain samples $u(t_l)$ and $y(t_l)$. This is achieved by rewriting the fitting error~\eqref{eq:vf_eVF} in the time-domain, where multiplication by partial fraction $1/(s-p_n)$ becomes a convolution between $e^{p_n t}$ and the input or output samples. These convolutions can be computed by numerical integrations, leading to a time-domain version of the original VF algorithm which closely follows the steps of the original frequency-domain VF algorithm~\cite{grivet2015passive}. 

The time-domain VF algorithm leads to a model in the continuous-time domain. Alternatively, if the sampling period $\Delta t = t_{l+1} - t_l$ is constant, one can also apply the z-domain VF~\cite{mekonnen2007broadband}, which relies on the z transform as opposed to the Laplace transform. This latter algorithm leads to a model in the discrete-time domain, which can be expressed as a digital filter or as a set of difference equations (as opposed to differential equations).

\subsection{Improved Vector Fitting formulations}

In the QuadVF algorithm~\cite{drmac2015quadrature}, a quadrature rule inspired by the $H_2$ error measure is used in conjunction with a suitable choice of frequency sampling points to improve the fidelity of the reduced model to the given samples. The same work also shows how one can incorporate derivative information, making QuadVF able to minimize a discrete Sobolev norm. In~\cite{drmac2015vector}, this approach is extended to the multi-input multi-output case, and a way to control the McMillan degree\footnote{The McMillan degree~\cite{zhou1996robust} of a matrix transfer function $H(s)$ is the order of a minimal state space realization of $H(s)$, such as the order $N$ of the Gilbert realization discussed in Sec.~\ref{sec:vf_realization}.} of the approximation is proposed, which helps to achieve smaller reduced models when $\bar{q}$ and $\bar{m}$ are high.

The numerical robustness of VF, which is already quite remarkable in its original formulation, is further improved in the Orthonormal VF algorithm~\cite{deschrijver2007orthonormal}. This algorithm replaces partial fractions $1/(s-p_n)$ in~\eqref{eq:vf_w} and~\eqref{eq:vf_H} with orthonormal rational functions, achieving better numerical conditioning of the linear system~\eqref{eq:vf_sysmimo} to be solved. 

Another subject that received considerable attention is the robustness of VF against noise in the given samples $H_k$. Noise may arise from the measurement process or, if samples were obtained with a numerical simulation, from round-off errors, approximations, and convergence issues. The relaxed normalization discussed in Sec.~\ref{sec:vf_relaxed} improves VF convergence in presence of noise~\cite{gustavsen2006improving}. Furthermore, the VF with adding and skimming includes a mechanism to detect spurious poles caused by noise~\cite{grivet2006improving} . Since spurious poles impair VF convergence, they must be removed throughout iterations~\cite{grivet2006improving}. This mechanism is coupled with a robust way to adaptively refine model order $\bar{n}$ to maximize accuracy even when noise is significant~\cite{grivet2006improving}. Taking into account noise variance in the definition of the VF fitting error was also shown to improve convergence~\cite{ferranti2010variance}. Finally, instrumental variables can be used to unbias the VF process from the effects of noise, leading to better accuracy and convergence at no additional cost~\cite{beygi2012instrumental}.

\subsection{VF algorithms for distributed systems}

The efficient modeling of distributed systems is an open problem in model order reduction. A system is distributed when its size is not negligible compared to wavelength. Hence, the time an acoustic or electromagnetic wave takes to propagate through the system cannot be neglected. Propagation delays lead to the presence of irrational terms in the transfer function of the underlying system. Typically, these terms are in the form $e^{-s \tau}$ where $\tau$ is the propagation delay. Rational functions, including the partial fractions in~\eqref{eq:vf_Hmimofinal} can accurately fit these irrational terms, up to arbitrary 
accuracy. However, if $\tau$ is not negligible, the required order may be large, and will quickly increase as $\tau$ grows. This leads to a large model which may burden subsequent simulations.

To overcome this issue, the core idea is to explicitly include exponential terms $e^{-s \tau_l}$ in the reduced model which will be fitted to the given samples. A popular choice is to define each element $\widetilde{H}_{qm}(s)$ of the model transfer function as
\begin{equation}
	\sum_{l=1}^{\bar{l}} \left( r_{0,l} +  \sum_{n=1}^{\bar{n}_l} \frac{r_{n,l}}{s-p_{n,l}}   \right) e^{-s \tau_l}\,,
	\label{eq:vf_Hdistrib}
\end{equation} 
where the $_{qm}$ subscript was omitted from all coefficients for clarity. The exponential factors in~\eqref{eq:vf_Hdistrib} are meant to efficiently capture long propagation delays, while the rational terms between brackets will resolve the residual behavior of the system. Typically, since long propagation delays are already accounted for by the exponential terms, the order $\bar{n}_l$ of the rational factors can be kept quite low.

For systems with uniform cross-section along the direction of propagation, such as electrical transmission lines, VF is used in conjunction to the method of characteristics to obtain an efficient distributed model~\cite{kuznetsov1996optimal,achar2001simulation,grivet2004transient,morched1999universal}. For distributed systems of general shape, several VF algorithms with delay terms have been proposed~\cite{chinea2008compact,chinea2010delay,charest2010time,triverio2010identification,roy2011transient,luo2012extended}. In these algorithms, the first step is to identify the values of the relevant propagation delays $\tau_i$ present in the system. Given only frequency samples $H_k$, this is not a trivial task, and the dominant approach is to exploit time-frequency decompositions~\cite{gustavsen2004time,grivet2006delay,roy2011transient,kocar2016new}. Next, the coefficients of the remaining rational factors in the model are determined with a VF-like iterative process~\cite{chinea2010delay,charest2010time,triverio2010identification,roy2011transient,luo2012extended}.

\subsection{Parametric VF algorithms}

The design process of an engineering system typically requires a large number of simulations for different values of design parameters, such as material properties, geometrical dimensions and operating conditions (e.g. bias voltages, temperature,...). In early design stages, parametric simulations are used to explore the design space. Later on, they may be used to optimize design in order to meet specifications or improve performance. Moreover, parametric simulations also help designers to account for manufacturing variability during design. In the context of parametric simulations, conventional VF models may be inefficient. Indeed, every time a parameter changes, a new set of samples $H_k$ must be obtained, and the fitting process has to be repeated from scratch. 

A better solution is to create a parametric VF model which captures the system response with respect to both frequency $s$ and some parameters of interest $\mu^{(1)}, \mu^{(2)},...$. The core idea behind parametric VF techniques~\cite{triverio2007parametric,triverio2008improved,jnl-2010-temc-macromodeling4emi,deschrijver2008robust,grivet2017compact} is to let residues $R_n$ and poles $p_n$ in~\eqref{eq:vf_Hmimofinal} be parameter-dependent functions, such as polynomials in $\mu^{(1)}, \mu^{(2)},...$. Their coefficients can be determined with an iterative process analogous to the Sanathanan-Koerner iteration in Sec.~\ref{sec:vf_sk}, starting from samples of the system's frequency response obtained for multiple values of parameters $\mu^{(1)}, \mu^{(2)},...$. The main advantage of a parametric model is that, once generated, it can be reused many times for different parameter values within its range of validity. One of the challenges in the generation of parametric VF models is how to guarantee that the model will be stable and passive over the desired parameter range~\cite{triverio2009parameterized,triverio2010passive,ferranti2010guaranteed}. Recently, systematic solutions to this challenging problem have been proposed~\cite{zanco2018enforcing}.

\section{Conclusion}

This chapter introduced the Vector Fitting algorithm, which has become one of the most popular tools for the extraction of linear reduced-order models from samples of their response, collected in the frequency or in the time domain. Vector Fitting produces a rational model which approximately minimizes the least-squares error between the given samples and the model response. Determining model coefficients is originally a nonlinear least-squares problem, whose solution is prone to the typical issues of nonlinear minimization: high computational cost and problematic convergence due to local minima. Vector Fitting overcomes these issues by iteratively minimizing a linearization of the original problem, leveraging well-established methods for the solution of linear least-squares problems. Several strategies to obtain a robust and efficient implementation of VF have been reviewed. When properly implemented, Vector Fitting enjoys remarkable robustness, efficiency and versatility, typically converging in a handful of iterations. Finally, we reviewed the most prominent extensions of the original algorithm which have been proposed for data-driven modeling of time-domain systems, noisy samples, distributed systems, and parametric systems.

Vector Fitting's superior performance and reliability lead to a widespread use in many different fields. Originally conceived to predict how transients propagate throughout power distribution networks, VF is the method of choice for the wideband modeling of overhead lines, underground cables and power transformers~\cite{morched1999universal, gustavsen2004wide, noda2005identification, annakkage2012dynamic, grivet2015passive}. In electronic engineering, VF is extensively used to model the propagation of high-speed signals through interconnect networks found at the chip, package and printed circuit board level. These models are crucial for system design, and greatly help in preventing signal integrity, power integrity and electromagnetic compatibility issues~\cite{achar2001simulation, ruehli2001progress, li2010progress, swaminathan2010designing,jnl-2010-temc-macromodeling4emi, jnl-2011-tcmpt-channels, wu2013overview}. The impact of VF in this area is confirmed by the fact that all leading commercial tools for the design of high-frequency electronic circuits include a VF module.  Applications in microwave engineering~\cite{liao2007vector, triverio2010extraction,deschrijver2011microwave,de2012characterization} and digital filter design~\cite{wong2008iir} have also been reported. Within computational electromagnetism, VF can be used to efficiently model the Green's function of layered media, which is necessary to solve Maxwell's equations with integral equation methods~\cite{kourkoulos2006accurate, boix2007application, polimeridis2007robust}. The ability of VF to generate models compatible with transient simulations has also been exploited in the finite difference time domain (FDTD) method~\cite{lin2012fdtd,michalski2013low}, the finite element time domain method~\cite{cai2009faster,wang2009incorporation}, and the discontinuous Galerkin method~\cite{yan2018analysis}. Beyond electrical engineering, VF found countless applications in various domains, including acoustics~\cite{cotte2009time,robinson2013characterizing}, fluid dynamics~\cite{almondo2006time,jaensch2016robust,grivet2015passive},  mechanical engineering~\cite{grivet2015passive,balmes1997garteur} and in the thermal modeling of chemical batteries~\cite{hu2012state}. For a collection of VF applications and additional references, the Reader is referred to~\cite{grivet2015passive}.


\bibliographystyle{abbrv}
\bibliography{vf_chapter}

\end{document}